\documentclass[11pt]{article}
\usepackage{fullpage}
\usepackage{graphicx}
\usepackage{upref}
\usepackage{enumerate}
\usepackage{latexsym}
\usepackage{pdfpages}
\usepackage[bookmarks]{hyperref}
\usepackage{color,graphics}
\usepackage{comment} 
\usepackage{caption}
\usepackage{subcaption}
\usepackage{wrapfig}
\usepackage{braket}

\usepackage{amssymb}
\usepackage{amsmath}
\usepackage{bbm}
\usepackage{amsthm}
\usepackage{mathrsfs}
\usepackage{mathtools}

\usepackage{epsfig,graphicx}
\usepackage{algorithmic}
\usepackage[ruled,linesnumbered]{algorithm2e}
\usepackage{amsfonts}
\usepackage{tikz} 
\usepackage{varioref}
\usepackage[ansinew]{inputenc}
\usepackage{qcircuit}
\usepackage{authblk}
\usepackage{multirow}
\usepackage{lineno}

\theoremstyle{plain}
\newtheorem{theorem}{Theorem}
\theoremstyle{plain}
\newtheorem{lemma}[theorem]{Lemma}
\theoremstyle{plain}

\newtheorem{corollary}[theorem]{Corollary}
\theoremstyle{plain}

\theoremstyle{plain}

\theoremstyle{plain}

\theoremstyle{definition}

\theoremstyle{definition}

\theoremstyle{remark}
\newtheorem{remark}{Remark}[section]

\theoremstyle{definition}

\AtBeginDocument{}

\newcommand{\id}{\mathbb{I}}

\newcommand{\had}{\text{H}}
\newcommand{\T}{\text{T}}

\newcommand{\conj}[1]{\overline{#1}}
\newcommand{\indx}{\mathcal{I}}
\newcommand{\bin}{\text{bin}}
\newcommand{\tcount}{\mathcal{T}}
\newcommand{\tofcount}{\mathcal{T}^{of}}
\newcommand{\qubit}{\mathcal{Q}}
\newcommand{\cnotcount}{\mathcal{C}}

\newcommand{\qram}{\text{QRAM}}
\newcommand{\qlut}{\text{qLUT}}

\begin{document}

\title{A quantum random access memory (QRAM) using a polynomial encoding of binary strings}  
\author{Priyanka Mukhopadhyay }

\affil[*]{ mukhopadhyay.priyanka@gmail.com, priyanka.mukhopadhyay@utoronto.ca  }

\affil[1]{Department of Computer Science, University of Toronto, ON, Canada}

\date{}

\maketitle

\begin{abstract}
 Quantum algorithms claim significant speedup over their classical counterparts for solving many problems. An important aspect of many of these algorithms is the existence of a quantum oracle, which needs to be implemented efficiently in order to realize the claimed advantages in practice. A quantum random access memory (QRAM) is a promising architecture for realizing these oracles. In this paper we develop a new design for QRAM and implement it with Clifford+T circuit. We focus on optimizing the T-count and T-depth since non-Clifford gates are the most expensive to implement fault-tolerantly in most error correction schemes. Integral to our design is a polynomial encoding of bit strings and so we refer to this design as $\qram_{poly}$. Compared to the previous state-of-the-art bucket brigade architecture for QRAM, we achieve an exponential improvement in T-depth, while reducing T-count and keeping the qubit-count same. Specifically, if $N$ is the number of memory locations to be queried, then $\qram_{poly}$ has T-depth $O(\log\log N)$, T-count $O(N-\log N)$ and uses $O(N)$ logical qubits, while the bucket brigade circuit has T-depth $O(\log N)$, T-count $O(N)$ and uses $O(N)$ qubits. Combining two $\qram_{poly}$ we design a quantum look-up-table, $\qlut_{poly}$, that has T-depth $O(\log\log N)$, T-count $O(\sqrt{N})$ and qubit count $O(\sqrt{N})$. A quantum look-up table (qLUT) or quantum read-only memory (QROM) has restricted functionality than a QRAM. For example, it cannot write into a memory location and the circuit needs to be compiled each time the contents of the memory change. The previous state-of-the-art CSWAP architecture has T-depth $O(\sqrt{N})$, T-count $O(\sqrt{N})$ and qubit count $O(\sqrt{N})$. Thus we achieve a double exponential improvement in T-depth while keeping the T-count and qubit-count asymptotically same. Additionally, with our polynomial encoding of bit strings, we develop a method to optimize the Toffoli-count of circuits, specially those consisting of multi-controlled-NOT gates.
\end{abstract}


\section{Introduction}
\label{sec:intro}

Quantum computers hold immense promise to outperform classical computers in many applications. Over the years numerous quantum algorithms have been developed that claim speedups over their classical counterparts in various problems, for example, unstructured database search \cite{1996_G}, optimization \cite{2020_AGGW}, quantum chemistry algorithms \cite{2017_RWSetal, 2018_BWMetal, 2020_BBMetal, 2019_CROetal, 2023_RBMetal}, data processing for machine learning \cite{2009_O,2017_BWPetal, 2009_HHL, 2022_HKTetal, 2018_CHIetal, 2015_AADetal, 2019_BDLK}, cryptography \cite{1994_S, 2013_K}, image processing \cite{2021_OCKK}. Many of these algorithms require access to oracles in order to fetch classical data and in practice, this is a non-trivial task. It is important to specify the details of implementations of these oracles in order to claim a genuine quantum speedup \cite{2015_A}. Efficient implementation of oracles can reduce the crossover of runtime between classical and quantum advantage from years to days \cite{2021_BLHetal}.

Till date, the most general-purpose design for the implementation of quantum oracles is a quantum random access memory (QRAM) \cite{2008_GLM, 2005_KMR, 2010_B, 2023_LWSetal, 2023_PCG}, which analogous to a classical random access memory (RAM), returns the element stored in a particular memory location. Specifically, suppose there are $N$ memory locations, each indexed by an integer $i\in\{0,1,\ldots,N-1\}$ and element $x_i$ is stored in location $i$. Then on input $i$, a classical RAM returns $x_i$. This procedure is called ``reading'' from the memory. A classical RAM is also able to ``write'' a particular data $x_i$ into memory location $i$. With a QRAM we are able to query a superposition of addresses. Let $A$ be the input qubit register storing the memory address to be queried and $B$ be the output register. If $\ket{\psi_{in}}$ and $\ket{\psi_{out}}$ are the input and output states, then
\begin{eqnarray}
 \ket{\psi_{in}} &=& \sum_{i=0}^{N-1} \alpha_i\ket{i}^A\ket{0}^B    \nonumber \\
 \ket{\psi_{out}} &=& \sum_{i=0}^{N-1} \alpha_i\ket{i}^A\ket{x_i}^B.
 \label{eqn:inOutRead}
\end{eqnarray}
The above equations correspond to the process equivalent to ``reading''. Like its classical counterpart, a QRAM is also able to ``write'' into a memory location. In this case, the input state is
\begin{eqnarray}
 \ket{\psi_{in}'} &=& \sum_{i=0}^{N-1} \alpha_i\ket{i}^A\ket{x_i}^B,   
 \label{eqn:inWrite}
\end{eqnarray}
and after the operation $x_i$ is XOR-ed into the memory location $i$. The oracles described in many algorithms \cite{2017_KP} do require the writing operation. Giovannetti, Lloyd and Maccone introduced the fanout and bucket-brigade architectures for QRAM in their pioneering work in \cite{2008_GLM, 2008_GLM_prl}. Since then much work has been done to study these designs and improve upon them. Out of the two desgins the bucket-brigade QRAM has become the most popular because it has better noise resilience \cite{2015_AGJetal, 2021_HLGetal} and fault-tolerant resource estimates \cite{2020_dMGM}. Several proposals for the experimental implementations of QRAM have been put forth \cite{2008_GLM, 2012_HXZetal, 2016_MM, 2019_HZZetal, 2021_CDEetal, 2022_P, 2024_WPG}, each utilizing the bucket brigade architecture. In \cite{2023_XHFetal} the authors propose a design implementing an $n$-bit QRAM on hardware nominally supported only on an $m$-bit query, where $m<n$. Over the years there have been proposals for various implementations of QRAM using different techniques, often for specific applications \cite{2016_AMGetal, 2018_SRWM, 2019_JS, 2019_PPR, 2020_VAPS, 2021_ASW, 2021_OCKK, 2021_ZAKetal, 2022_DSB, 2023_ASW, 2023_BCCetal, 2023_CAY, 2023_CXWetal, 2023_PLG, 2023_LHJ, 2024_LJ, 2024_DH, 2024_H}, and thus QRAMs have been used for a wide variety of tasks like neural networks, data processing, quantum communication, image processing, cryptanalysis, quantum simulation, circuit synthesis, state preparation, etc.

A circuit implementing a QRAM needs to be compiled and optimized only once, while the contents of the memory are free to change. But it has the disadvantage of a significant space overhead. A bucket-brigade QRAM for $N$ memory locations require $O(N)$ T gates, $O(N)$ ancillae and has T-depth $O(\log N)$ \cite{2020_dMGM}. In order to reduce the number of ancillae many algorithms use a sequence of multi-controlled-NOT gates, also known as quantum read-only memory (QROM) \cite{2018_BWMetal, 2020_dMGM, 2023_RBMetal, 2024_MSW}. This can be implemented with $O(N)$ T gates, $O(\log N)$ ancillae and $O(N)$ T-depth. Inspite of a lower qubit count, one disadvantage of a QROM is the exponentially higher T-depth which is not desirable for an efficient fault-tolerant implementation. Another disadvantage of QROM is the fact that we need to know the contents of the memory in advance. Each time the database changes, the circuit needs to be recompiled and optimized. There are other architectures, as in \cite{2019_PPR, 2020_VAPS}, that perform queries in $O(N\log N)$ time using $O(\log N)$ qubits.

Many hybrid architectures have been proposed that interpolate between these two extremes and leverage their space-time tradeoff \cite{2019_BGMetal, 2020_dMGM, 2022_HKRS, 2022_KSRZ, 2024_LKS, 2024_ZSL, 2024_MSW}. Notable among these is the CSWAP architecture \cite{2024_LKS}, which can be thought of as a combination of a QROM and a specific QRAM. It has T-count $O(\sqrt{N})$, number of ancillae $O(\sqrt{N})$ and T-depth $O(\sqrt{N})$. Here we mention that in literature the QRAM, QROM and these hybrid architectures are also used to build quantum look-up-table (qLUT) and so the names are often used interchangeably. These are required to perform restricted tasks. The contents of the table or database are known and this can be leveraged to design circuits with better resource estimates like T-count.

\subsection{Our contributions}
\label{subsec:contribution}

In this paper we propose an architecture for a QRAM, mainly aimed at reducing the non-Clifford gate complexity of the circuit implementation. We implement our circuits with the fault-tolerant, universal Clifford+T gate set because it implements more unitaries exactly compared to other fault-tolerant, universal gate sets \cite{2021_MM, 2024_Mcs, 2024_Mtof,2024_Mv}. In most error correction schemes the cost of implementing the non-Clifford T gate is significantly higher than the Clifford gates. Thus it is important to optimize the number of T-gates or T-count. It is also important to optimize the T-depth \cite{2012_F, 2012_FMMC, 2016_AMGetal, 2020_HS, 2020_dMGM, 2022_GMM}, which is related to the running time. A T-depth-1 corresponds to a set of T gates that can be implemented in parallel. We also refer to it as a layer or stage. So a T-depth $\tcount_d$ for a circuit implies $\tcount_d$ such stages, where in each stage the T gates are implemented in parallel. The product of T-depth and number of logical qubits is taken as a parameter to measure the rough cost of fault-tolerant implementation in the surface code \cite{2016_AMGetal, 2020_dMGM, 2024_LKS}. Our contributions in this paper can be summarized in the following points.

\begin{table}[h]
 \centering
 \begin{tabular}{|c|c|c|c|}
  \hline
  & T-depth & T-count & $\#$Logical qubits \\
  \hline
  Bucket-brigade \cite{2008_GLM, 2020_dMGM} & $O(\log N)$ & $O(N)$ & $O(N)$    \\
  \hline
  CSWAP \cite{2024_LKS} & $O(\sqrt{N})$ & $O(\sqrt{N})$ & $O(\sqrt{N})$ \\ \hline
  $\qram_{poly}$ (This work) & $O(\log\log N)$ & $O(N)$ & $O(N)$  \\
  \hline
  $\qlut_{poly}$ (This work) & $O(\log\log N)$ & $O(\sqrt{N})$ & $O(\sqrt{N})$  \\
  \hline
 \end{tabular}
\caption{Comparison of T-depth, T-count and number of logical qubits required to implement QRAM and qLUT.}
\label{tab:compare}
\end{table}

(I) We develop a quantum random access memory, which we call $\qram_{poly}$ (Section \ref{sec:qram}), with the help of a polynomial encoding of bit strings (Section \ref{sec:polyEncoding}). We show that $\qram_{poly}$ can be implemented with $N-\log N-1$ Toffoli gates. We can parallelize this circuit, using an additional $O(N)$ ancillae in order to achieve a Toffoli-depth of $\log \log N$. This implies a T-count of $O(N-\log N-1)$ and T-depth of $O(\log \log N)$. Thus compared to previous bucket-brigade architecture \cite{2020_dMGM} we achieve an exponential improvement in the T-depth, reduce the T-count, while keeping the number of ancillae nearly the same. 

(II) We use two $\qram_{poly}$ to design a QROM or qLUT. This is a hybrid architecture and we achieve a T-count of $O(\sqrt{N})$, T-depth $O(\log\log N)$ and ancillae count of $O(\sqrt{N})$. Thus, here we achieve a double exponential improvement in T-depth compared to previous designs \cite{2024_LKS, 2024_ZSL}, while keeping the T-count and ancillae count asymptotically similar. We refer to this desgin as $\qlut_{poly}$ (Section \ref{sec:qLUT}). In Table \ref{tab:compare} we have summarized and compared the cost of implementation of our $\qram_{poly}$ and $\qlut_{poly}$ with some previous works.

(III) The encoding polynomials that are integral to our constructions of QRAM and qLUT have other applications. For example, in Section \ref{subsec:TcountOpt} we describe a procedure (TOFFOLI-OPT-POLY) to optimize the Toffoli-count of circuits. Later we also discuss some other potential applications and hence these polynomials may be of independent interest.

\subsection{Organization}

In Section \ref{sec:polyEncoding} we describe a polynomial encoding of bit strings. Using this we design our $\qram_{poly}$ in Section \ref{sec:qram}. The design of $\qlut_{poly}$ and a method for circuit optimization has been discussed in Section \ref{sec:application}. Finally we conclude in Section \ref{sec:discuss}.

\section{Polynomial encoding of Boolean strings}
\label{sec:polyEncoding}

In this section we describe an encoding where a bit string of length $n$ is represented by a polynomial and then we derive certain properties of the set of $N=2^n$ polynomials. These attributes will aid in the design of $\qram_{poly}$, as explained in later sections.

\paragraph{Notations : } We use the following notations and conventions. A \textbf{polynomial} in $n$ variables comprises of a sum of one or more \textbf{monomials}, where each monomial is the product of at most $n$ variables. We say that a monomial has \textbf{weight} $k$ if it is the product of $k$ variables. A \textbf{constant} is a monomial of weight 0. A polynomial is \textbf{linear} if it can be expressed as sum of monomials of weight at most 1.
Let $\indx\subseteq\{1,2,\ldots,n\}$ be a subset of indices of the variables $x_1,\ldots,x_n$. We refer to the subscripts as indices. We denote a monomial with variables having indices in $\indx$ by $m_{\indx}$. That is,
\begin{eqnarray}
 m_{\indx} = \prod_{j\in \indx} x_j.
 \label{eqn:monoDefn}
\end{eqnarray}

\paragraph{Encoding polynomial :} Suppose we have an $n$-length bit string - ($b_1,b_2,\ldots, b_n$), denoted as $\vec{b}$. We encode this bit string into a polynomial in $n$ Boolean variables - $x_1, x_2,\ldots, x_n$, where variable $x_i$ corresponds to bit $b_i$. We assign the following polynomial to each variable $b_i$.
\begin{eqnarray}
 b_i &\mapsto& \frac{1+(-1)^{b_i}}{2} + x_i := p_{b_i}(x_i)
 \label{eqn:bitEncode}
\end{eqnarray}
If $b_i = 0$ then $b_i\mapsto \frac{1+1}{2}+x_i = 1+x_i$ and if $b_i = 1$ then $b_i\mapsto\frac{1-1}{2}+x_i = x_i$. The complete bit string $(b_1,b_2,\ldots, b_n)$ is encoded as follows.
\begin{eqnarray}
 (b_1,b_2,\ldots, b_n) \mapsto \prod_{i=1}^n\left(\frac{1+(-1)^{b_i}}{2}+x_i \right) = \prod_{i=1}^np_{b_i}(x_i) := p_{\vec{b}}(x_1,\ldots,x_n) 
 \label{eqn:stringEncode}
\end{eqnarray}
Now we prove some properties of the encoding polynomials.
\begin{lemma}
Suppose we have $n$ bits - $b_1,b_2,\ldots,b_n$ and we associate a variable $x_i$ to each bit $b_i$. Consider the $2^n$ encoding polynomials $\{ p_{\vec{b}}(x_1,\ldots,x_n) \}$ corresponding to the $2^n$ possible $n$-bit strings $\vec{b} = (b_1, b_2,\ldots,b_n)$, as defined in Equation \ref{eqn:stringEncode}. Then we have
\begin{eqnarray}
 p_{\vec{b}}(b_1',b_2',\ldots,b_n') \equiv \delta_{\vec{b},\vec{b'}} \mod 2,\qquad\text{where}\quad \vec{b'} = (b_1', b_2',\ldots,b_n'), \nonumber
\end{eqnarray}
implying $p_{\vec{b}}\left(b_1',b_2',\ldots,b_n'\right)\equiv 1\mod 2$ if and only if $\vec{b} = \vec{b'}$ or $b_j = b_j'$ for each $j=1,\ldots,n$. Else, it is $0 \mod 2$.
 \label{lem:uniqueVal}
\end{lemma}

\begin{proof}
 By definition of the encoding in Equation \ref{eqn:bitEncode}, $p_{b_i}(x_i) = 1+x_i$ when $b_i = 0$ and $p_{b_i}(x_i) = x_i$ when $b_i=1$. Thus, $p_{b_i}(b_i) = 1$ and since $p_{\vec{b}}(x_1,\ldots,x_n) = \prod_{i=1}^np_{b_i}(x_i)  $, so $p_{\vec{b}}(b_1,\ldots,b_n)\equiv 1\mod 2$. 
 
 Again, if $b_i' \neq b_i$ then $p_{b_i}(b_i') = 2$ or $0$. So $p_{\vec{b}}(b_1',\ldots,b_n')\equiv 0\mod 2$, whenever $\vec{b'} \neq \vec{b}$.
 
This proves the lemma.
\end{proof}

\begin{lemma}
Let $p_{\vec{b}}(x_1,\ldots,x_n)$ be the encoding polynomial corresponding to the bit string $\vec{b} = (b_1,\ldots,b_n)$, as defined in Equation \ref{eqn:stringEncode}. Assume that $k$ of the bits i.e. $b_{i_1},\ldots,b_{i_k}$ are 1 and the rest 0. Then, 
\begin{eqnarray}
 p_{\vec{b}}(x_1,\ldots,x_n) = \sum_{\mathcal{I}': \mathcal{I}'\supseteq \mathcal{I} } m_{\mathcal{I}'}, \qquad\text{where}\qquad\mathcal{I} = \{ i_1,\ldots, i_k\}.   \nonumber
\end{eqnarray}
 \label{lem:monoSum}
\end{lemma}

\begin{proof}
  Let $\conj{\mathcal{I}} = \{1,\ldots,n\}\setminus \mathcal{I}$ be the complement set of $\mathcal{I}$. All additions and multiplications are commutative. By definition,
 \begin{eqnarray}
  p_{\vec{b}}(x_1,\ldots,x_n) &=& \left(\prod_{j:j\in \mathcal{I}}x_j\right)\left(\prod_{ \ell:\ell\in \conj{\mathcal{I}}} \left(1+x_{\ell}\right) \right)   \nonumber \\
  &=&\left(\prod_{j:j\in \mathcal{I} }x_j\right)\left(1+\sum_{\ell\in \conj{\mathcal{I}}} x_{\ell}+\sum_{\substack{ \ell_1\neq \ell_2 \\ \ell_1,\ell_2\in\conj{\mathcal{I}} }} x_{\ell_1}x_{\ell_2}+\cdots+\prod_{\ell\in\conj{\mathcal{I}}} x_{\ell}  \right),
 \end{eqnarray}
which clearly proves the lemma.
\end{proof}

We have the following corollaries.  
\begin{corollary}
Let $p_{\vec{b}}(x_1,\ldots,x_n)$ be the encoding polynomial corresponding to the bit string $\vec{b} = (b_1,\ldots, b_n)$, as defined in Equation \ref{eqn:stringEncode}. Let $\mathcal{I}_1$ be the subset of indices of the bits in $\vec{b}$ that have value 1. Given any subset of indices $\mathcal{I}\subseteq \{1,\ldots,n\}$, the monomial $m_{\mathcal{I}}$ appears as a summand in $p_{\vec{b}}(x_1,\ldots,x_n)$ if and only if $\mathcal{I}_1 \subseteq \mathcal{I}$. 
\label{corr:monoAppear}
\end{corollary}

\begin{corollary}
 \begin{eqnarray}
  \prod_{j=1}^n\left(1+x_j\right)=1+\sum_{j=1}^kx_j+\sum_{j\neq k}x_jx_k+\sum_{j\neq k\neq \ell}x_jx_kx_{\ell}+\cdots+\prod_{j=1}^nx_j.
  \nonumber
 \end{eqnarray}
\label{corr:fullExpand}
 \end{corollary}

\begin{corollary}
Each encoding polynomial $p_{\vec{b}}(x_1,\ldots,x_n)$, defined in Equation \ref{eqn:stringEncode}, has exactly one summand monomial of minimum weight. Specifically, let $\mathcal{I}_1$ be the subset of indices of the bits in $\vec{b}$ that have value 1. Then the minimum weight monomial is
\begin{eqnarray}
 m_{\mathcal{I}_1} = \prod_{j\in \mathcal{I}_1} x_j. \nonumber
\end{eqnarray}

Also, it follows that each encoding polynomial has a unique minimum weight monomial.
\label{corr:uniqueMinMono}
\end{corollary}

\textbf{New labeling :} Thus we can label each encoding polynomial $p_{\vec{b}}(x_1,\ldots,x_n)$ by $p_{m_{\mathcal{I}}}(x_1,\ldots,x_n)$, where $m_{\mathcal{I}} = \prod_{j\in\mathcal{I}}x_j$ is the minimum weight monomial and $\mathcal{I}\subseteq\{1,\ldots,n\}$ is the subset of indices of the bits in $\vec{b}$ that have value 1.

\begin{theorem}
Let $p_{\vec{b}}(x_1,\ldots,x_n)$ be the encoding polynomial corresponding to a bit string $\vec{b}=(b_1,\ldots,b_n)$, as defined in Equation \ref{eqn:stringEncode}. Suppose $\mathcal{I}_1\subseteq\{1,\ldots,n\}$ is the subset of indices of the bits in $\vec{b}$ that have value 1. Then,
\begin{eqnarray}
 p_{\vec{b}}(x_1,\ldots, x_n) = p_{m_{\mathcal{I}_1}} (x_1,\ldots,x_n) = m_{\mathcal{I}_1}+ \bigoplus_{\mathcal{I}':\mathcal{I}_1\subset\mathcal{I}'} p_{m_{\mathcal{I}'}}. \nonumber
\end{eqnarray}
In the above by XOR we mean that the coefficients of same monomials are added modulo 2.
 \label{thm:monoXOR}
\end{theorem}

\begin{proof}
 From Corollary \ref{corr:uniqueMinMono}, $m_{\mathcal{I}_1}$ appears as a summand in $p_{\vec{b}}(x_1,\ldots,x_n)$. Let $|\mathcal{I}_1|=w$. We need to prove that any monomial $m_{\mathcal{I}'}$ such that $\mathcal{I}'\supset\mathcal{I}_1$ will be added odd number of times.
 
 Consider the sets $\mathcal{I}_{2j} = \mathcal{I}_1\bigcup\{j\}$ such that $j\notin\mathcal{I}_1$. If we add polynomials $p_{m_{\mathcal{I}_{2j}}}$ which have $m_{\indx_{2j}}$ as the minimum weight monomial then each of these monomials of weight $w+1$ gets added only once.  
 
 Consider the sets $\indx_{3jk}=\indx_1\bigcup\{j,k\}$ such that $j,k\notin\indx_1$. Now for each pair of indices $j,k\notin\indx_1$ we have $\indx_{3jk} = \indx_{2j}\bigcup\{k\} = \indx_{2k}\bigcup\{j\}$. From Lemma \ref{lem:monoSum}, the monomial $m_{\indx_{3jk}}$ appears as a summand in $p_{m_{\indx_{2j}}}$ and $p_{m_{\indx_{2k}}}$. It also appears as the minimum weight monomial in $p_{m_{\indx_{3jk}}}$. From Corollary \ref{corr:monoAppear} it cannot appear in any other polynomial $p_{m_{\indx'}}$ where $\indx'\supset\indx_1$. Thus monomials of the form $m_{\indx_{3jk}}$ with weight $w+2$ gets added odd number of times. 
 
 Similarly we can generalize this argument to monomials of weight $w'>w$.  
 Consider the index set $\indx_{w'} =\indx_1\bigcup\indx_{2}$ such that $|\indx_2|=w'-w$. Thus the monomial $m_{\indx_{w'}}$ has weight $w'$. From Lemma \ref{lem:monoSum} and Corollary \ref{corr:monoAppear} this monomial appears as a summand in all polynomials of the form $p_{m_{\indx''}}$, where 
 \begin{eqnarray}
 \indx_1\subset\indx''\subseteq\indx_{w'}.  \label{eqn:thm:subsetRel}
 \end{eqnarray}
 Number of subsets of weight $w+\ell$ such that they satisfy the subset relation in Equation \ref{eqn:thm:subsetRel} is $\binom{w'-w}{\ell}$. Here $\ell$ varies from $1$ to $w'-w$. Thus number of times $m_{\indx_{w'}}$ gets added is
 \begin{eqnarray}
  \sum_{\ell=1}^{w'-w}\binom{w'-w}{\ell} = \sum_{\ell=0}^{w'-w}\binom{w'-w}{\ell} -1 = 2^{w'-w}-1 \equiv 1\mod 2. \nonumber
 \end{eqnarray}

This proves the theorem.

\end{proof}

In Tables \ref{tab:encodePoly3} and \ref{tab:encodePoly4} we have enlisted all the encoding polynomials for 3 and 4-bit strings. We have also specified the alternate labeling of each polynomial, that is, indexed by its unique minimum weight monomial. The various properties proved in this section can be verified from these tables.

\begin{table}[h]
 \centering
 \begin{tabular}{|c|c|c|}
 \hline
 $\vec{\mathbf{b}}$ & $\mathbf{p}_{\vec{\mathbf{b}}} \mathbf{(x_1,x_2,x_3)}$ &  \\
  \hline
  000 & $1+x_1+x_2+x_3+x_1x_2+x_2x_3+x_1x_3+x_1x_2x_3$ & $p_1$ \\
  \hline
  001 & $x_3+x_1x_3+x_2x_3+x_1x_2x_3$ & $p_{x_3}$ \\
  \hline
  010 & $x_2+x_1x_2+x_2x_3+x_1x_2x_3$ & $p_{x_2}$ \\
  \hline
  011 & $x_2x_3+x_1x_2x_3$ & $p_{x_2x_3}$ \\
  \hline
  100 & $x_1+x_1x_2+x_1x_3+x_1x_2x_3$ & $p_{x_1}$ \\
  \hline
  101 & $x_1x_3+x_1x_2x_3$ & $p_{x_1x_3}$ \\
  \hline
  110 & $x_1x_2+x_1x_2x_3$ & $p_{x_1x_2}$ \\
  \hline
  111 & $x_1x_2x_3$ & $p_{x_1x_2x_3}$ \\
  \hline
 \end{tabular}
\caption{Encoding polynomials for 3 bit strings. In the last column an alternate labeling has been mentioned where each polynomial is indexed by its unique minimum weight monomial.}
\label{tab:encodePoly3}
\end{table}

\begin{table}[h]
 \centering
 \scriptsize
 \begin{tabular}{|c|p{15cm}|c|}
 \hline
 $\vec{\mathbf{b}}$ & $\mathbf{p}_{\vec{\mathbf{b}}} \mathbf{(x_1,x_2,x_3,x_4)}$ &  \\
  \hline
  0000 & $1+x_1+x_2+x_3++x_4+x_1x_2+x_1x_3+x_1x_4+x_2x_3+x_2x_4+x_3x_4+x_1x_2x_3+x_1x_2x_4+x_1x_3x_4+x_2x_3x_4+x_1x_2x_3x_4$ & $p_1$ \\
  \hline
  0001 & $x_4+x_1x_4+x_2x_4+x_3x_4+x_1x_2x_4+x_1x_3x_4+x_2x_3x_4+x_1x_2x_3x_4$ & $p_{x_4}$ \\
  \hline
  0010 & $x_3+x_1x_3+x_2x_3+x_3x_4+x_1x_2x_3+x_1x_3x_4+x_2x_3x_4+x_1x_2x_3x_4$ & $p_{x_3}$ \\
  \hline
  0011 & $x_3x_4+x_1x_3x_4+x_2x_3x_4+x_1x_2x_3x_4$ & $p_{x_3x_4}$ \\
  \hline
  0100 & $x_2+x_1x_2+x_2x_3+x_2x_4+x_1x_2x_3+x_1x_2x_4+x_2x_3x_4+x_1x_2x_3x_4$ & $p_{x_2}$ \\
  \hline
  0101 & $x_2x_4+x_1x_2x_4+x_2x_3x_4+x_1x_2x_3x_4$ & $p_{x_2x_4}$ \\
  \hline
  0110 & $x_2x_3+x_1x_2x_3+x_2x_3x_4+x_1x_2x_3x_4$ & $p_{x_2x_3}$ \\
  \hline
  0111 & $x_2x_3x_4+x_1x_2x_3x_4$ & $p_{x_2x_3x_4}$ \\
  \hline
  1000 & $x_1+x_1x_2+x_1x_3+x_1x_4+x_1x_2x_3+x_1x_2x_4+x_1x_3x_4+x_1x_2x_3x_4$ & $p_{x_1}$ \\
  \hline
  1001 & $x_1x_4+x_1x_2x_4+x_1x_3x_4+x_1x_2x_3x_4$ & $p_{x_1x_4}$ \\
  \hline
  1010 & $x_1x_3+x_1x_2x_3+x_1x_3x_4+x_1x_2x_3x_4$ & $p_{x_1x_3}$ \\
  \hline
  1011 & $x_1x_3x_4+x_1x_2x_3x_4$ & $p_{x_1x_3x_4}$ \\
  \hline
  1100 & $x_1x_2+x_1x_2x_3+x_1x_2x_4+x_1x_2x_3x_4$ & $p_{x_1x_2}$ \\
  \hline
  1101 & $x_1x_2x_4+x_1x_2x_3x_4$ & $p_{x_1x_2x_4}$ \\
  \hline
  1110 & $x_1x_2x_3+x_1x_2x_3x_4$ & $p_{x_1x_2x_3}$ \\
  \hline
  1111 & $x_1x_2x_3x_4$ & $p_{x_1x_2x_3x_4}$ \\
  \hline
 \end{tabular}
\caption{Encoding polynomials for 4 bit strings. In the last column an alternate labeling has been mentioned where each polynomial is indexed by its unique minimum weight monomial.}
\label{tab:encodePoly4}
\end{table}

\section{Quantum random access memory ($\qram_{poly}$)}
\label{sec:qram}

In this section we describe the construction of $\qram_{poly}$ using the polynomial encoding of bit strings, discussed in the previous Section \ref{sec:polyEncoding}. We implement the circuits using Clifford+T gate set, as discussed earlier in Section \ref{subsec:contribution}. 

\paragraph{Some definitions :} Here we briefly recap the following definitions. The \textbf{T-count} of a circuit is the number of T-gates in it. The \textbf{Toffoli-count} of a circuit is the number of Toffoli-gates in it. Let $U$ be the unitary implemented by a circuit. Assume $U$ can be expressed as a product of $\tcount_d$ unitaries, i.e. $U = \prod_{j=1}^{\tcount_d}U_j$, where each $U_j$ is such that the T or $\T^{\dagger}$ gates appearing in its circuit can be implemented in parallel. We call $\tcount_d$ as the \textbf{T-depth} of the circuit for $U$. Each $U_j$ has T-depth 1 circuit. We can define the \textbf{Toffoli-depth} of a circuit in an analogous manner.

Suppose we have $N = 2^n$ memory locations, each specified or indexed by an $n$-bit string $\vec{b} = (b_1,\ldots,b_n)$, which is its \textbf{address}. We have $n$ input qubits $\{q_1,\ldots,q_n\}$, whose state selects a memory location. We call these \textbf{address qubits}. The main difference between a qubit and a bit is the fact that the former can be in a superposition of both the $\ket{0}$ and $\ket{1}$ states, while the latter can either have state (or value) 0 or 1. Thus state of qubits $(q_1,\ldots,q_n)$ can be a superposition of bit strings $(b_1,\ldots, b_n)$, each specifying a particular memory location, say $M_{\vec{b}}$. We can encode each bit string with the encoding polynomial (Equation \ref{eqn:stringEncode}) described in Section \ref{sec:polyEncoding}. It follows that each memory location $M_{\vec{b}}$ is associated with a polynomial $p_{\vec{b}}(x_1,\ldots,x_n)$ in $n$ variables $\{x_1,\ldots,x_n\}$, uniquely determined by $\vec{b}$. We can alternatively call this its \textbf{polynomial address}. 

We first describe how one particular memory location with address $\vec{b} = (b_1,\ldots,b_n)$ is queried. That is, the state of the address qubit $q_j = b_j$, for each $j=1,\ldots,n$. Then it is straightforward to understand the operation of the $\qram_{poly}$ circuit, when a superposition of memory locations are queried. An illustration of 3-qubit $\qram_{poly}$ has been shown in Figure \ref{fig:qram3}.

\begin{figure}
\centering
\begin{subfigure}[b]{0.35\textwidth}
 \centering 
 \includegraphics[width=4cm, height=8cm]{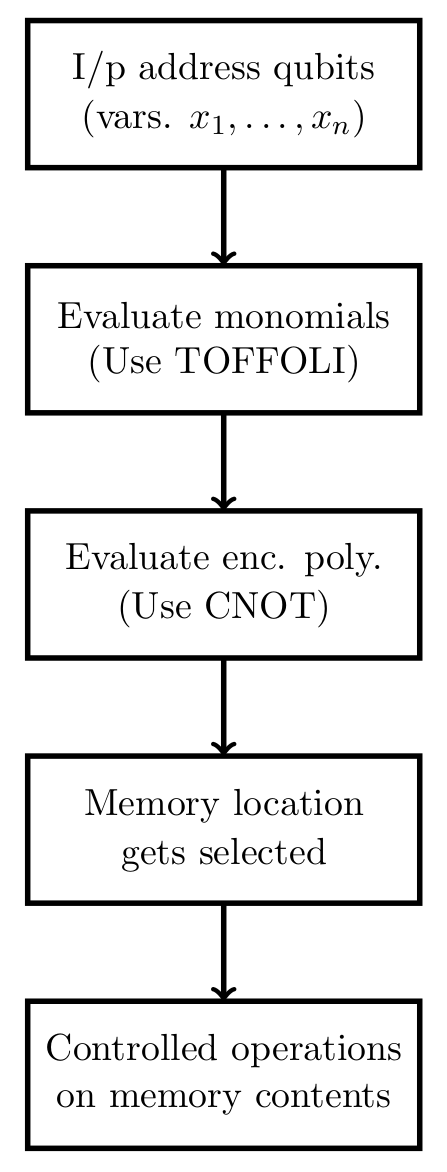}
 \caption{}
 \label{fig:flowchart}
\end{subfigure}
\hfill
\begin{subfigure}[b]{0.6\textwidth}
 \centering 
 \includegraphics[width=6cm, height=6cm]{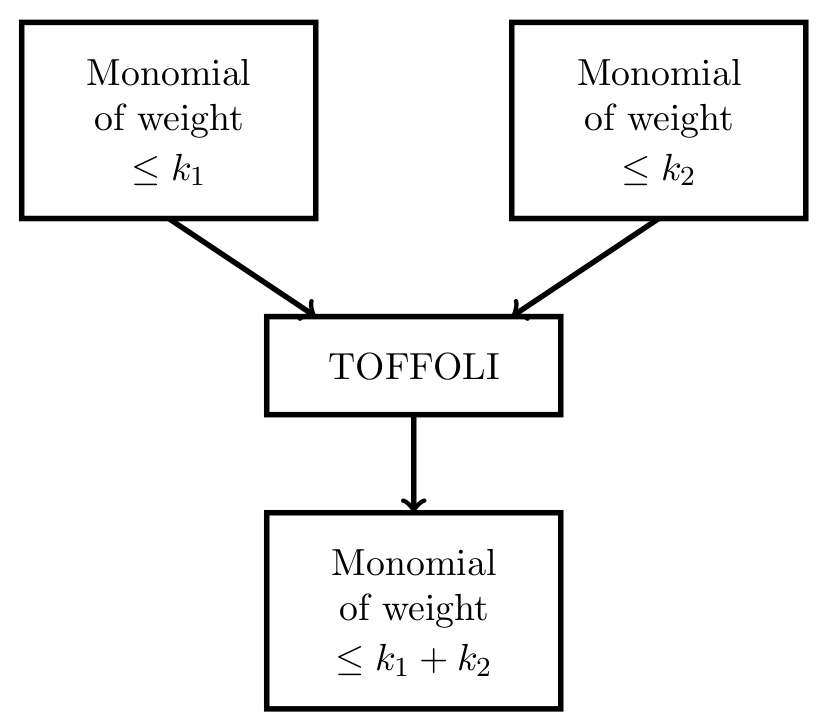}
 \caption{} 
 \label{fig:evalMono}
\end{subfigure}
 \caption{(a) A flowchart showing the procedures in $\qram_{poly}$. In the figure ``I/p'' and ``enc. poly.'' represents ``Input'' and ``encoding polynomials'', respectively.  (b) The procedure of evaluating monomials using Toffoli.  }
 \label{fig:flowMono}
\end{figure}

\begin{figure}
\centering
\begin{subfigure}[b]{0.6\textwidth}
\centering
 \includegraphics[width=10cm, height=5cm]{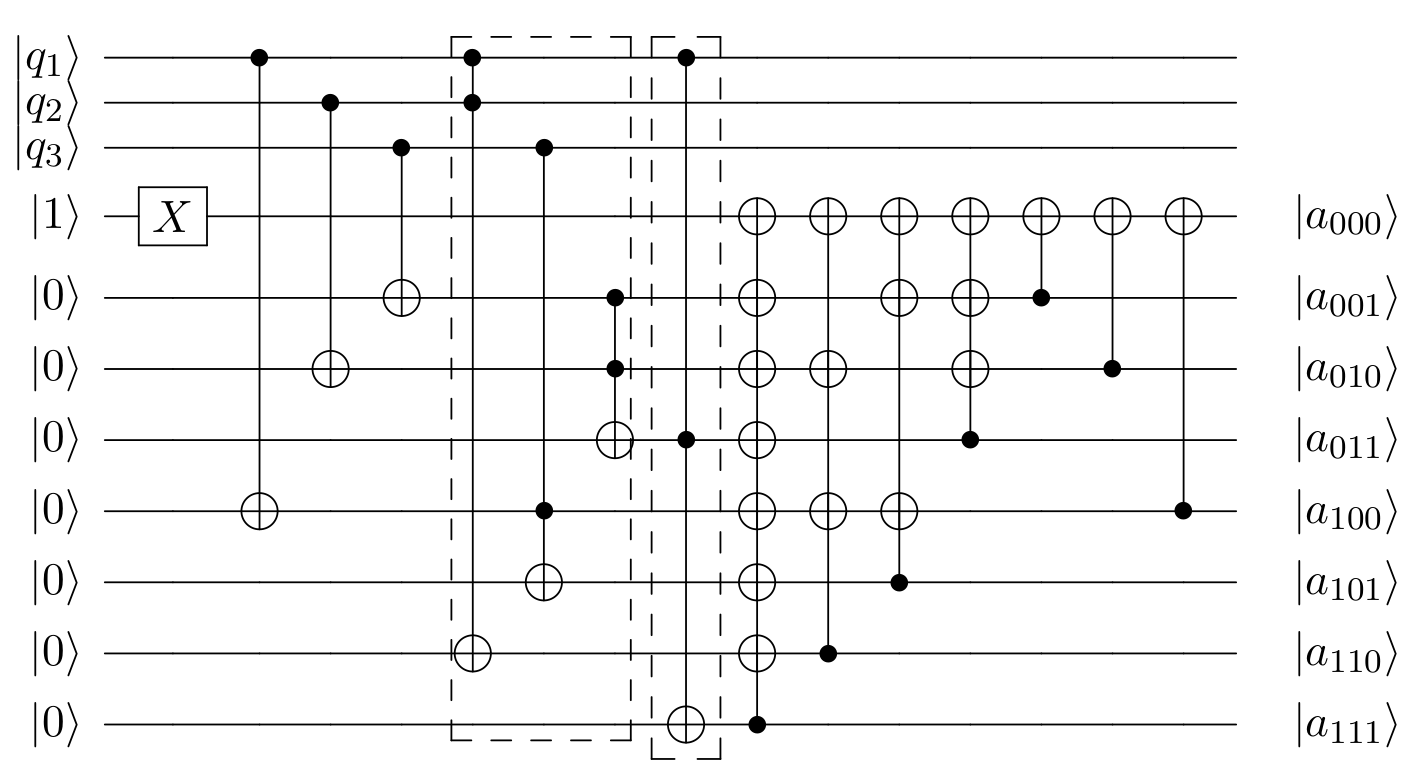}
 \caption{}
 \label{fig:qram3}
\end{subfigure}
\hfill
\begin{subfigure}[b]{0.3\textwidth}
\centering
 \includegraphics[width=\textwidth, height=\textwidth]{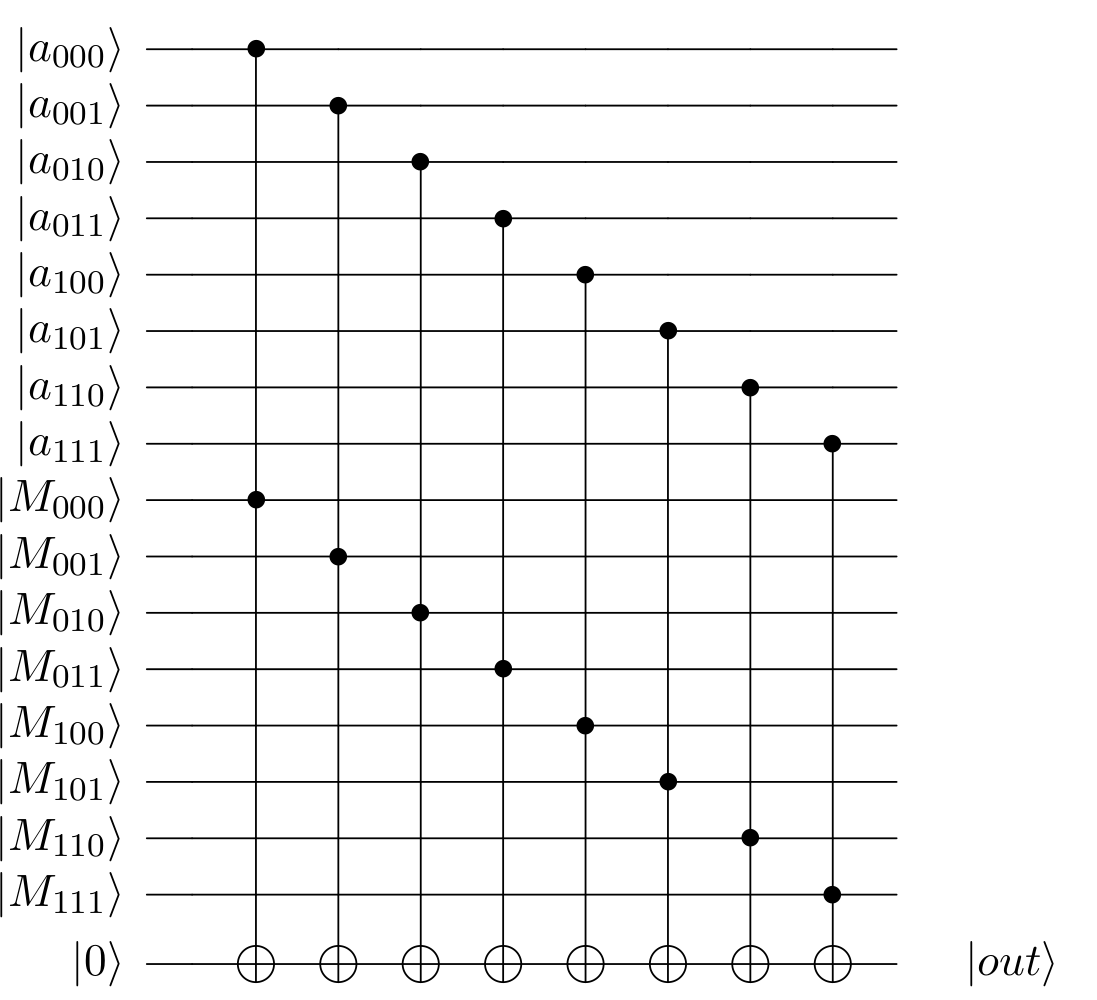}
 \caption{}
 \label{fig:parity}
\end{subfigure}
\hfill
 \begin{subfigure}[b]{0.4\textwidth}
 \includegraphics[width=8cm, height=7cm]{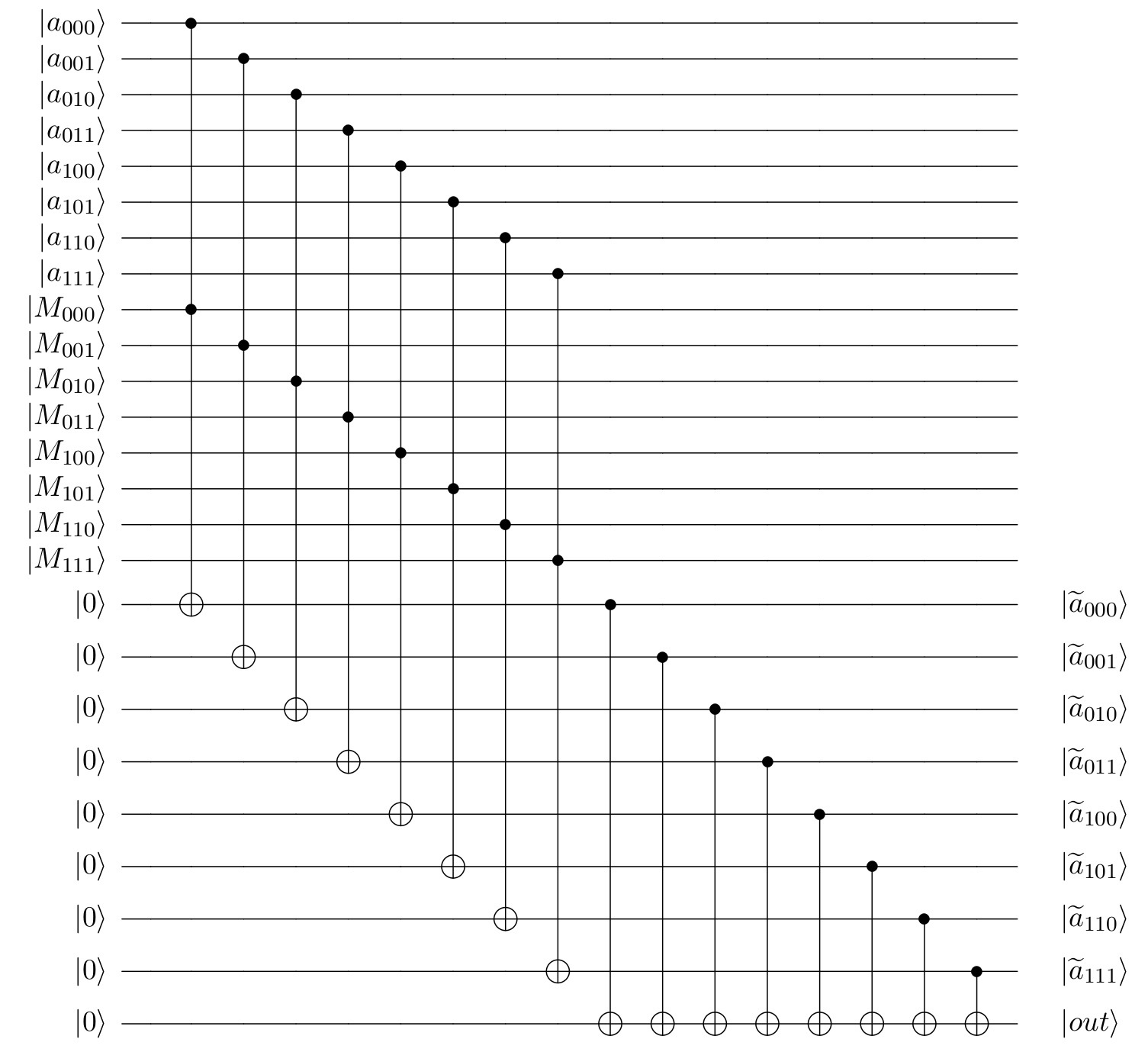}
 \caption{}
 \label{fig:Mpar}
\end{subfigure}
\caption{$\qram_{poly}$ on 3 qubits. (a) The circuit shows the computation of the encoding polynomials and their storage in specific ancilla. These ancillae are used to select memory locations for further operations like reading, writing, assigning phase, etc. Each dotted box corresponds to Toffoli-depth 1. Thus the circuit has Toffoli-depth 2. (b) The circuit shows the reading operation where contents of one of the memory location, controlled on an ancilla, is copied to the output bus. (c) A parallelized version of the circuit in (b). The Toffoli-depth in this case is 1.  }
\label{fig:memory3}
\end{figure}

We allocate $N$ ancillae $a_0, \ldots,a_{N-1}$, such that $a_j$ implements the encoding polynomial of $j$ (in the binary form). Each of these ancilla are initialized in state $\ket{0}$. Here we mention that for the remaining part of the paper we use either integers or their binary representation for indexing. This is for convenience and it should be clear from the context. From Lemma \ref{lem:uniqueVal} we know that for each of the $N$ possible bit strings only one of these ancilla flips to $\ket{1}$ and it is uniquely determined by the bit string. Thus these ancillae can be used to select memory locations. Each of the input qubits is assigned a variable. A Toffoli can be used to multiply two monomials because it operates as follows, with input states $\ket{x}, \ket{y}, \ket{0}$.
\begin{eqnarray}
 TOFFOLI\ket{x}\ket{y}\ket{0}\mapsto\ket{x}\ket{y}\ket{xy}
 \label{eqn:tofEqn}
\end{eqnarray}
We perform the following steps. 
 In Figure \ref{fig:flowMono} we have shown a flowchart depicting these steps.

\begin{enumerate}
 \item[\textbf{Step 1.}] \textbf{Computing monomials :} We implement the monomials using CNOTs and Toffolis. 
 This can be done by multiplying lower weight monomials, using Toffolis. Each monomial is stored in a specific ancilla. Suppose an ancilla $a_j$ is intended to select memory location $M_j$. Let $m_{\indx}$ is the minimum weight monomial of the encoding polynomial corresponding to the binary representation of $j$. Then $m_{\indx}$ is stored in $a_j$. By Corollary \ref{corr:uniqueMinMono}, each monomial gets stored in distinct and uniquely determined ancilla.
 
 \item[\textbf{Step 2.}] \textbf{Computing encoding polynomials :} Using CNOTs we XOR the monomials and implement the encoding polynomials, as stated in Theorem \ref{thm:monoXOR}. That is, after this step ancilla $a_j$ stores $p_{\vec{b}}(x_1,\ldots,x_n)$, where $\vec{b}$ is the binary representation of $j$. So it can be used to select the memory location $M_j$.
 
 \item[\textbf{Step 3.}] \textbf{Select and compute :} Using a Toffoli controlled on $a_j$ and $M_j$ we copy the memory content onto the output bus (Figure \ref{fig:parity}). This is equivalent to ``reading'' from the memory. For ``writing'' into the memory we reverse the control and target at the output bus and $M_j$. 
 
 \item[\textbf{Step 4.}] \textbf{Making the operations reversible :} To obtain a fully reversible $\qram_{poly}$, after the reading or writing operation we implement the circuit, as described in steps 1 and 2, in the reverse order. This is equivalent to uncomputation.
\end{enumerate}

As an example, consider the 3-qubit $\qram_{poly}$ shown in Figure \ref{fig:memory3}. Apart from the 3 input qubits $q_1, q_2, q_3$ there are $N=2^3=8$ ancillae (Figure \ref{fig:qram3}), each intended to select a memory location. We label these ancillae as $a_{000}, a_{001},\ldots,a_{111}$ and the corresponding memory locations as $M_{000}, M_{001},\ldots, M_{111}$, respectively. For simplicity, we assume each memory location has 1 qubit. Now in each ancilla $a_{\vec{b}}$ we want to implement the polynomial $p_{\vec{b}}$ (Table \ref{tab:encodePoly3}), which is a sum of monomials in variables $x_1, x_2, x_3$, assigned to input qubits $q_1, q_2, q_3$, respectively. First, we use 3 CNOTs to get the 3 minimum weight monomials of weight 1, that is, $x_1, x_2, x_3$ and store them in $a_{100}, a_{010}, a_{001}$, respectively. Next, using Toffolis we compute monomials of weight 2, that is, $x_1x_2, x_2x_3, x_1x_3$ and store these in ancillae $a_{110}, a_{011}, a_{101}$, respectively. After that we again use Toffoli to multiply monomials of weight 1 and 2, obtaining $x_1x_2x_3$, and store it in $a_{111}$. We observe (refer Table \ref{tab:encodePoly3}) that if a monomial $m_{\indx}$ is the minimum weight monomial of an encoding polynomial $p_{\vec{b}}$ then it is stored in ancilla $a_{\vec{b}}$. This completes Step 1.

Then, using CNOTs we XOR the monomials and obtain the encoding polynomials in corresponding ancilla, as stated in Theorem \ref{thm:monoXOR}. This implies we first compute encoding polynomials whose minimum weight monomial has highest weight, then we compute those polynomials whose minimum weight monomial has the second highest weight and so on. One way of doing this step is to XOR $a_{\indx}$ (storing $m_{\indx}$) with each $a_{\indx'}$ (storing $m_{\indx'}$) such that $\indx'\subset\indx$. We start from $\indx$ with highest cardinality $n$, then sets of indices with second highest cardinality $n-1$ and so on. For example, in Figure \ref{fig:qram3} the highest weight monomial is $x_1x_2x_3$ and after step 1, $p_{x_1x_2x_3}$ is already computed in $a_{111}$. We XOR $a_{111}$ with all other ancillae because $\{1,2,3\}$ is a superset of the set of indices of all other monomials. This also computes the polynomials $p_{x_1x_2}$, $p_{x_2x_3}$ and $p_{x_1x_3}$ in $a_{110}$, $a_{011}$ and $a_{101}$, respectively. Next, we XOR $a_{\indx}$ with each $a_{\indx'}$ such that $|\indx|=2$ and $\indx'\subset\indx$. So, $a_{110}$ is XORed with $a_{100}$, $a_{010}$ and $a_{000}$. Similarly, we XOR $a_{011}$ and $a_{101}$ with 3 other ancillae (each). This completes the computation of polynomials $p_{x_1}$, $p_{x_2}$ and $p_{x_3}$ in $a_{100}$, $a_{010}$ and $a_{001}$, respectively. Finally, we XOR $a_{100}$, $a_{010}$ and $a_{001}$ with $a_{000}$. We use X gate on $a_{000}$ to add 1. This completes the computation of polynomial $p_{1}$. This also completes Step 2. 

In Step 3 if we want to read from the memory then we use the circuit in Figure \ref{fig:parity}. Here controlled on each ancilla a memory location is copied to the output bus $\ket{out}$.


\subsection{Illustration : Application of $\qram_{poly}$ in Grover's algorithm}
\label{subsec:grover}

To further illustrate the application of $\qram_{poly}$ we consider the Grover's algorithm \cite{1996_G}, which gives a thoeretical quadratic speedup over classical search algorithms in an unstructured database. Suppose there are $N$ items in the database. The Grover's algorithm consists of the following steps. First we initialize the system in the state $\ket{\Psi} = \ket{0}^n$. Then we perform a number of iterations of the following procedure.
\begin{enumerate}
 \item  Set all qubits into an equal superposition state $\ket{s}$.
\begin{eqnarray}
 \had^{\otimes n}\ket{0}^n &=&\frac{1}{\sqrt{N}} \sum_{i=0}^{N-1}\ket{i} = \ket{s}. \nonumber
\end{eqnarray}

\item Phase-tag the states that represent the values to be searched.

\item Implement a diffusion operator $U_d = 2\ket{s}\bra{s}-\id$ that amplifies the amplitudes for measuring the states that need to be searched. 
\end{enumerate}
At the end of all iterations of the algorithm we perform a measurement in the computational basis. The searched items can be be found by identifying the distinct peaks in the distribution of the measured results. 

Steps 2 (phase-tagging) and 3 (diffusion operator) correspond to two successively performed reflections, and thus together they perform a rotation in a 2D-plane. Thus in each iteration of the Grover's algorithm the state $\ket{s}$ is rotated closer to a state $\ket{k}$, that represents a value to be searched. After an optimal number of iterations $\ket{s}$ is rotated the closest to $\ket{k}$. Searching one item in an unstructured database with $N$ items requires at most $O\left(\sqrt{N}\right)$ iterations. Classically this search can be done in $O(N)$ time complexity. Thus it is possible to achieve a quadatic speed-up, provided each iteration is done efficiently, or simply put, time complexity of each iteration is considerably less than the number of iterations. 

Usually, in theoretical analyses of Grover's algorithm we assume the existence of a phase-tagging oracle that performs step 2. This oracle has the following functionality :
\begin{eqnarray}
 O\ket{i} = (-1)^{f(i)}\ket{i}\quad\text{with}\quad f(i) = 1 \quad\text{if}\quad i\in\{k\}\quad\text{else}\quad f(i) = 0.  \nonumber
\end{eqnarray}
An efficient implementation of this oracle is essential in order to achieve the claimed speedup of the Grover's algorithm in practice. Our $\qram_{poly}$ can be used to implement this oracle. Specifically, after we compute the encoding polynomials in order to select a memory location (Step 3) we do not require the Toffolis, as shown in Figure \ref{fig:parity}. Instead, we use CZ on each memory location where the control is on the selecting ancilla. In this way, we apply phase on selected memory locations.

We are not going into more detail of an optimal fault-tolerant implementation of Grover's algorithm in order to achieve a practical quantum speed-up as this is a stand-alone research topic \cite{2023_SBBetal} and beyond the scope of this paper. But briefly we want to summarize this section by emphasizing that an efficient implementation of the phase-tagging oracle is crucial for the practical speed-up of Grover's algorithm. A QRAM (with proper modification) can be used and faster this QRAM, the better it is. In later sections we give a detail analysis of the cost of fault-tolerant implementation of $\qram_{poly}$ (though with read/write operations) with the surface code and show that it is much faster and consumes less number of qubits, compared to other QRAMs. Thus it can also be used to have a faster fault-tolerant implementation of Grover's algorithm.

\subsection{Cost of implementation}
\label{subsec:costCkt}

Now we discuss the cost of implementing a circuit with the Clifford+T gate set. We focus on optimizing the non-Clifford resource, that is, the T-count and T-depth, as discussed earlier in Section \ref{subsec:contribution}. We first bound the Toffoli depth and number of compute-uncompute Toffoli pairs. In literature there are different implementations of a Toffoli gate with the Clifford+T gate set, some optimizing its T-count, while there are others that optimize T-depth. The circuit in \cite{2013_J, 2018_G} gives the lowest T-count of 4 and has a T-depth of 2. It uses logical AND gadget which does the multiplication. One advantage of this circuit is the fact that if we have a compute-uncompute pair then the uncomputation part does not require any T-gate, but it uses classical measurements. Another circuit is the one in \cite{2013_S} which has a T-count of 7, T-depth of 1 and uses 4 extra ancillae. Depending on whichever parameter we want to optimize, one implementation can be favoured over the other.

We have separated the cost after the computation of the encoding polynomials since the cost of this part can change depending upon the required operations, as discussed in Section \ref{subsec:grover}. 

\paragraph{Number of compute-uncompute Toffoli pairs : } From Theorem \ref{thm:monoXOR} we know that we need to implement all monomials of weight $0,1,2,\ldots,n$. We do not require any Toffoli to implement monomials of weight 0 and 1. The former can be obtained by applying X gate and the latter are the variables assigned to the input qubits. There are $\binom{n}{2}$ monomials of weight 2. We need these many Toffolis in order to get all monomials of weight 2. Monomials of weight 3 can be obtained by multiplying each weight 2 monomial with a variable. In general, a monomial of weight $k\geq 2$ can be obtained by multiplying any two already calculated monomials of weights $k_1, k_2<k$, such that $k_1+k_2 = k$. Thus we require 1 Toffoli to compute each monomial of weight more than 1. We also require equal number of Toffolis in order to uncompute. Hence the number of compute-uncompute pairs of Toffoli we require is
\begin{eqnarray}
\tofcount_c = \sum_{k = 2}^n\binom{n}{k} = \sum_{k=0}^n\binom{n}{k}-n-1 = 2^n-n-1 = N - \log_2N-1.   
 \label{eqn:tofcount_c}
\end{eqnarray}
Additionally, for copying memory contents (Figure \ref{fig:parity}), we require $N\ell$ Toffolis, where $\ell$ is the number of qubits in each memory location or its size. This implies that $\qram_{poly}$ has a T-count of $O(N-\log_2 N-1)$ for computing the encoding polynomials. It requires an additional $O(N\ell)$ T-gates for reading or writing.

\paragraph{Number of logical qubits : } Apart from the $n = \log_2N$ qubits containing the input address, we require $N$ ancillae in order to select memory locations. Thus the number of logical qubits, excluding the $N\ell$ memory qubits is
\begin{eqnarray}
 \qubit = N+\log_2N.    \label{eqn:qubit}
\end{eqnarray}

\paragraph{Number of CNOT pairs :} Suppose an ancilla $a_j$ selects memory location $M_j$. We have already discussed that we need to implement the encoding polynomial of the binary representation of $j$ in $a_j$. From Corollary \ref{corr:uniqueMinMono} we know that each encoding polynomial has a unique minimum weight monomial determined by the bit string that it encodes. We can label each ancilla using these minimum weight monomials. Let $\indx$ be the set of indices of the bits in the binary representation of $j$, that have value 1. $m_{\indx}$ is the corresponding minimum weight monomial of the encoding polynomial $p_{\bin(j)}(x_1,\ldots,x_n)$, where $\bin(j)$ is the binary representation of $j$. Then we can alternatively refer to $a_j$ as $a_{\indx}$. 

Initially, using $n$ CNOTs we store the single weight monomials $x_1,\ldots,x_n$ in $n$ ancillae - $a_{\{1\}},\ldots,a_{\{n\}}$, respectively. Then using Toffolis we compute and store monomials of weight greater than 1 in different ancillae. According to Theorem \ref{thm:monoXOR} we use CNOTs to add these monomials in order to implement the encoding polynomials. Consider a monomial $m_{\indx}$ of weight $k$, i.e. $|\indx| = k$ that has been computed in ancilla $a_{\indx}$. From Corollary \ref{corr:monoAppear} we know that we need to add a CNOT from the ancilla  $a_{\indx}$ to each ancilla $a_{\indx'}$, where $\indx'\subset\indx$. Now $\indx$ has $2^k-1$ subsets (excluding itself). And there are $\binom{n}{k}$ monomials of weight $k$. We need an equal number of CNOT for uncomputations. Thus total number of CNOT pairs required is
\begin{eqnarray}
\cnotcount &=& n+\sum_{k=1}^n\binom{n}{k}(2^k-1) = n+\sum_{k=0}^n\binom{n}{k}2^k -\binom{n}{0}2^0 - \sum_{k=0}^n\binom{n}{k} +\binom{n}{0} = n+3^n-2^n \nonumber \\
 &=& \log_2N + N^{\log_23} - N \approx \log_2N+N^{1.6}-N.
 \label{eqn:numCNOT}
\end{eqnarray}

\paragraph{Toffoli-depth :} In this design Toffolis are required to compute the different monomials that are stored in distinct ancillae. As mentioned, we can compute a monomial of weight $k$ by multiplying two already-computed monomials of weight $k_1, k_2 < k$ such that $k_1+k_2 = k$. Initially we have $n$ monomials of weight 1 (the inputs) that are also stored in $n$ ancillae. So we have $2n$ monomials and using these we can compute $\frac{2n}{2} = n$ new monomials of higher weight in parallel. After that we can compute $\frac{2n+n}{2}=\frac{3n}{2}$ monomials in parallel using the already available $3n$ monomials. Next, we can compute $\frac{1}{2}\left(2n+\frac{2n}{2}+\frac{1}{2}\left(2n+\frac{2n}{2}\right)\right)$ monomials in parallel using the available monomials. Roughly, we can compute $O(n)$ monomials in parallel in each round. Since we need to compute $2^n-n-1$ monomials of weight greater than 1, so Toffoli-depth is 
\begin{eqnarray}
 \tofcount_d \in O\left(\frac{2^n-n-1}{n}\right)\in O\left(\frac{N-\log_2N-1}{\log_2N}\right).
 \label{eqn:Tdepth}
\end{eqnarray}

\subsection{Parallelizing the multiplications : lower T-depth}
\label{subsec:qramPar}

\begin{figure}[!t]
 \centering
 \includegraphics[width=13cm, height=8cm]{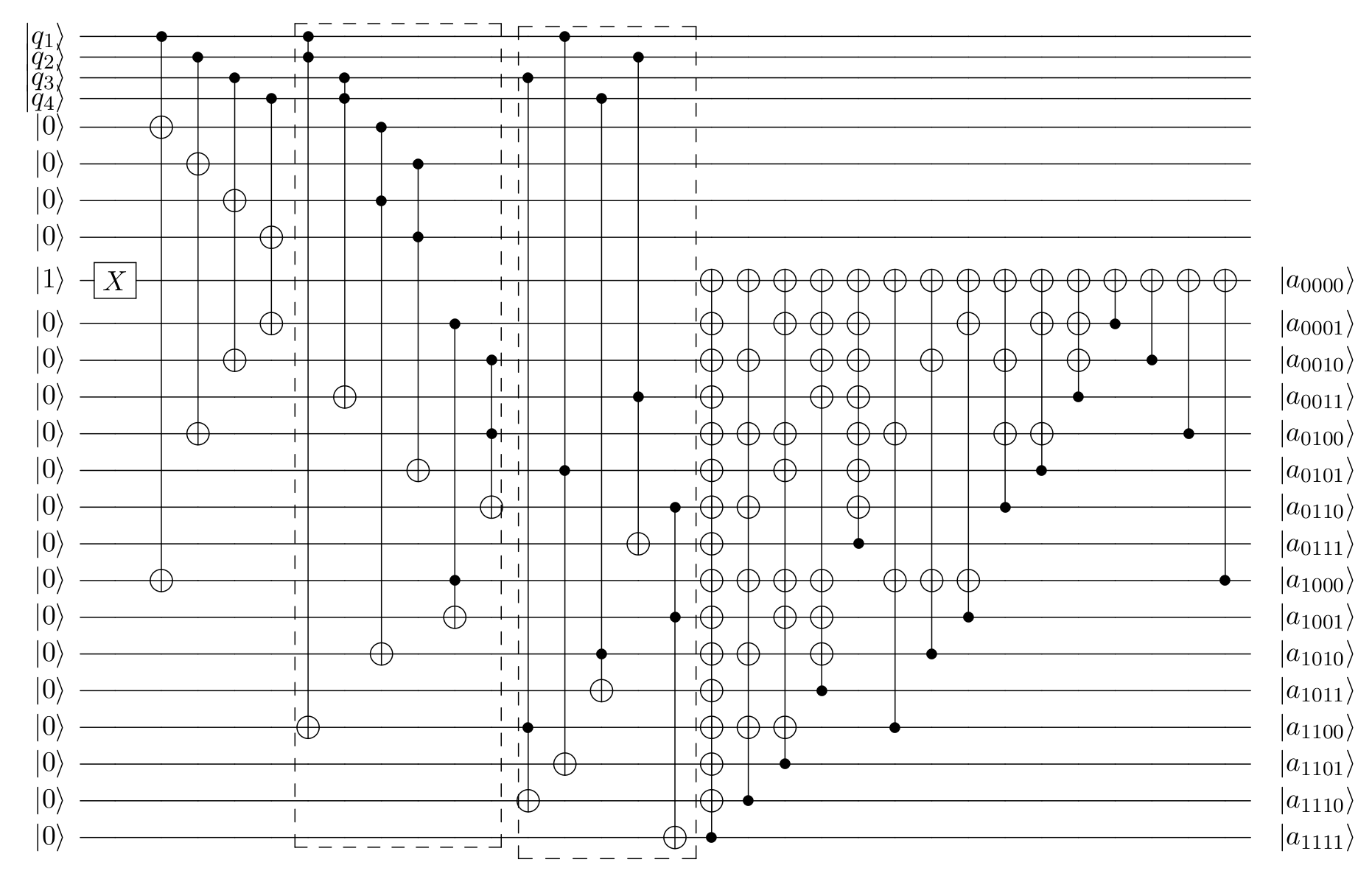}
 \caption{$\qram_{poly}$ on 4 qubits. This is a parallelized version, where the Toffoli-depth has been reduced by using extra ancillae. Each dotted box corresponds to Toffoli-depth 1. Thus the circuit has Toffoli-depth 2. We show only till the computation of the encoding polynomials in respective ancillae.}
 \label{fig:qram4}
\end{figure}

We observe that in this design the non-Clifford Toffolis are required to compute monomials of weight at most $n$. Since we store the monomials in different qubits so we can compute in parallel monomials of weight $k$ using monomials of weight $1,\ldots,\frac{k}{2}$. Thus the maximum weight of any monomial that can be computed in parallel is double the maximum weight of any monomial computed in the previous parallel stage (Each stage is Toffoli-depth 1). Hence, all the monomials can be computed with Toffoli-depth 
\begin{eqnarray}
 \tofcount_{d,par} = \log_2n = \log_2\log_2N.
 \label{eqn:TdepthAncilla}
\end{eqnarray} 
In Figure \ref{fig:qram4} we have shown a parallel implementation of a 4-qubit QRAM.

The set of $N\ell$ Toffolis required to read or write from the memory can be parallelized to have Toffoli-depth 1. A simple method using an additional $N\ell$ number of ancillae, has been shown in Figure \ref{fig:Mpar}, where $\ell=1$. During reading, the contents of each memory location is copied to an ancilla. Then using CNOTs the parity of these ancillae is stored on the output bus. As explained before, during writing the control and target of the memory locations and output bus are reversed. In order to parallelize these operations and have Toffoli-depth 1, we can also use the method described in Figure 4 of \cite{2020_dMGM} or in \cite{2020_POB}, the latter has exponentially less qubit-count.

Hence, using either of the implementations in \cite{2013_J, 2018_G} or \cite{2013_S}, accounting for both the computation and uncomputation part, we achieve a T-depth of $2\log_2\log_2N$, excluding the read/write operation.

\paragraph{Extra ancillae to reduce Toffoli-depth : } In the first step we compute monomials of weight 2, using monomials of weight 1. We store $n+n$ weight-1 monomials in $2n$ distinct qubits. Using these we can compute $n$ weight-2 monomials in parallel. We can compute the remaining $\binom{n}{2}-n$ weight-2 monomials by copying the inputs in different ancillae. So we require
\begin{eqnarray}
 2\left(\binom{n}{2} - n\right) \nonumber
\end{eqnarray}
extra ancillae. We reuse ancillae. In the next step (Toffoli-depth 2) we compute monomials of weight 3 and 4. We can already compute $\frac{1}{2}\left(2n+\binom{n}{2}\right)$ monomials using the monomials already stored (no extra ancilla). We can compute the remaining using
\begin{eqnarray}
 2\left(\binom{n}{4}+\binom{n}{3}\right) - \left(\binom{n}{2}+2n\right) \nonumber
\end{eqnarray}
extra ancillae. We use the ancillae used in Toffoli-depth-1. So we need not add this number to the previously calculated number. Generalizing, suppose we have computed all monomials of weight at most $k/2$. In the next step we can compute all monomials of weight $k$. We can compute $\frac{1}{2}\left(\binom{n}{k/2}+\cdots+2n\right)$ monomials using already stored monomials. Remaining can be computed using
\begin{eqnarray}
A_k =  2\left(\binom{n}{k}+\cdots+\binom{n}{k/2+1}\right)-\left(\binom{n}{k/2}+\cdots+\binom{n}{2}+2n\right)  \nonumber
\end{eqnarray}
extra ancillae. Number of extra ancillae we require is 
\begin{eqnarray}
 \max_kA_k \leq 2^n = N.    \label{eqn:extraAncTdepth}
\end{eqnarray}
Thus, excluding the memory qubits and ancillae required to parallelize the reading/writing operation, the total number of logical qubits we require is 
\begin{eqnarray}
\qubit_{par} \leq 2N +\log N. \label{eqn:numLogQubit2} 
\end{eqnarray}

\paragraph{Extra CNOT to reduce T-depth : } At each step we require CNOTs to copy monomials to extra ancillae and then again we reset. Roughly, we can say that we require $\max_kA_k$ CNOT pairs at each step. Thus number of extra CNOT pairs is at most
\begin{eqnarray}
 2^n\log_2n = N\log\log N.  \nonumber
\end{eqnarray}
So total number of CNOT pairs is 
\begin{eqnarray}
\cnotcount_{par} \in O\left(3^n-2^n+2^n\log_2n\right)\in O\left(N^{1.6}+N\log_2\log_2N\right).  \label{eqn:numCNOT2}
\end{eqnarray}

\begin{remark}
 A monomial of weight $k$ can be computed by $C^{k}X$, a NOT gate controlled on $k$ qubits. This gate, on input $\ket{x_1}, \ket{x_2},\ldots,\ket{x_k},\ket{0}$, returns the product $\ket{x_1x_2\ldots x_k}$, as follows.
 \begin{eqnarray}
  C^kX\ket{x_1}\ket{x_2}\cdots\ket{x_k}\ket{0}\mapsto\ket{x_1}\ket{x_2}\cdots\ket{x_k}\ket{x_1x_2\ldots x_k}    \nonumber
  \label{eqn:CnX}
 \end{eqnarray}
In principle, we can compute all the necessary $N-\log_2N-1$ monomials in parallel by using $\binom{n}{k}$ number of $C^kX$ gates, for each $k=2,\ldots,n$. We require enough number of ancillae in order to copy the input variables the required number of times. Thus for computing the encoding polynomials the T-depth or Toffoli-depth is determined by the maximum T-depth or Toffoli-depth required to implement any $C^kX$, where $k=2,\ldots,n$. So this part can improve with the design of better circuits for $C^nX$.
\label{remark:parMonomial}
\end{remark}

\subsection{Comparison with previous work}
\label{subsec:qramCompare}

We compare the resource estimates with the parallelized version of bucket-brigade QRAM \cite{2020_dMGM}. Here we mention that we have compared with the bucket-brigade architecture for the following reasons. First, this has been the most widely studied QRAM and a detailed fault-tolerant resource estimates is available, as in \cite{2020_dMGM}. Second, its applications are more general than other QRAMs, for example, FF-QRAM \cite{2019_PPR}, EQGAN-QRAM \cite{2022_NZBetal}, PQC-based QRAM \cite{2023_PLG}, which have been designed for specific problems and some of them do not scale well. Third, QRAMs as in \cite{2019_PPR, 2022_NZBetal, 2023_PLG, 2023_SBBetal, 2024_NZ} use unitaries like controlled rotations, that are approximately implementable by discrete universal gate sets. The non-Clifford cost like T-count or Toffoli-count varies inversely with the precision of synthesis \cite{2022_GMMapprox, 2024_Mtof, 2024_Mcs, 2024_Mv}. Hence, such designs are more expensive to implement fault-tolerantly. Fourth, many works have used the bucket-brigade QRAM as a basic module and designed more intricate architecture with it, most often to achieve some tradeoffs for particular applications or scenario, for example \cite{2023_XHFetal}. Thus, we can simply replace this module with our $\qram_{poly}$ in order to compare the designs. 

We compare the Toffoli-count, Toffoli-depth and number of logical qubits. In this way we do not have to worry about the difference in T-count and T-depth due to different implementations of Toffoli. We do not consider resource estimates for the reading/writing operation since this part can be implemented and optimized in a similar fashion and hence its cost can be regarded the same. Without this, the parallel bucket-brigade circuit in \cite{2020_dMGM} requires $N-2$ compute-uncompute Toffoli pairs, $2N+\log_2N$ logical qubits and has Toffoli-depth $\log_2N$. From Equations \ref{eqn:tofcount_c}, \ref{eqn:TdepthAncilla} and \ref{eqn:numLogQubit2}, this implies the following. 
\begin{eqnarray}
 \mathcal{R}_{T-depth} &=& \frac{\text{T-depth in this work}}{\text{T-depth in \cite{2020_dMGM}}} = \frac{\log_2\log_2N}{\log_2N} \label{eqn:ratioTdepth} \\
 \mathcal{R}_{T-count} &=& \frac{\text{T-count in this work}}{\text{T-count in \cite{2020_dMGM}}} = \frac{N-\log_2N-1}{N}   \label{eqn:ratioTcount} \\
 \mathcal{R}_{qubits} &=& \frac{\#\text{Logical qubits in this work}}{\#\text{Logical qubits in \cite{2020_dMGM}}} = \frac{2N}{2N} = 1  \label{eqn:ratioQubits}
\end{eqnarray}
Thus we achieve an exponential improvement of T-depth, reduction in T-count, while keeping the number of logical qubits the same.

\paragraph{Fault-tolerant implementation :} In most error-correction schemes, including the most popular surface code, the cost of implementation of the non-Clifford T gate is much more than the cost of the Clifford gates \cite{2012_FMMC}. In \cite{2020_dMGM} the following has been taken as a rough estimate for the cost of implementation with the surface code.
\begin{eqnarray}
 \text{Rough cost } = \log_2\left(\text{Logical qubits }\times\text{ T-depth}\right)
 \label{eqn:qVol}
\end{eqnarray}
Thus from Equations \ref{eqn:ratioTdepth}-\ref{eqn:ratioQubits} we can say that our $\qram_{poly}$ has much less fault-tolerant cost estimates from previous bucket-brigade architecture. So we expect much better performance in terms of running time.

For illustration, we consider an implementation of our $\qram_{poly}$ with the surface code, and estimate resource requirements following the procedures described in \cite{2012_FMMC} and \cite{2016_AMGetal}. Consider the number of qubits $n=36$, which corresponds to 8 GB of classical data. Using logical AND gadgets \cite{2018_G}, Toffoli can be implemented with 4 T gates with a T-depth 2 and if there is a compute-uncompute Toffoli pair we do not require any T gate for the uncomputation part. Hence, from Equation \ref{eqn:tofcount_c}, assuming $\ell=1$, the T-count is
\begin{eqnarray}
 \tcount \leq 4(2N-\log n -1) = 5.4975\times 10^{11}
\end{eqnarray}
and the T-depth is
\begin{eqnarray}
 \tcount_d \leq 2(\lceil\log n\rceil+1) = 14.
\end{eqnarray}
Each T gate requires one magic state ($\ket{A_L}$). For the above T-count the output error rate for state distillation should be no greater than
\begin{eqnarray}
 p_{out} = \frac{1}{\tcount}\approx 1.81899\time 10^{-12}. 
\end{eqnarray}
Now we use Algorithm 4 in \cite{2016_AMGetal} in order to calculate the distance of the surface code. Assuming a magic state injection error rate $p_{in} = 10^{-4}$, a per-gate error rate $p_g = 10^{-5}$, the stated algorithm suggests 2 layers of distillation with distances $d_1 = 10$, $d_2 = 5$. The first stage of distillation consumes $16$ logical qubits and the second stage consumes $16\times 15 = 240$ logical qubits in order to generate a single magic state. 

The input states of the logical qubits in the second layer are encoded on a distance $d_2 = 5$ code that uses $N_{p2} = 2.5\times 1.25\times d_2^2\approx 79$ physical qubits per logical qubit. The total footprint of the distillation circuit is then $240\times N_{p2} = 18750$ physical qubits. This round of distillation is completed in $\sigma_2 = 10d_2 = 50$ surface code cycles.

The first or top layer requires a $d_1 = 10$ surface code, for which a logical qubit takes $N_{p1} = 2.5\times 1.25\times d_1^2\approx 313$ physical qubits. So the total number of physical qubits required is $16\times N_{p1} = 5000$, with the round of distillation completed in $\sigma_1 = 10d_1 = 100$ surface code cycles.

The concatenated distillation scheme is performed in $\sigma = \sigma_1+\sigma_2=150$ surface code cycles. Since the first or top layer has lower footprint than the second or bottom layer, distillation can potentially be pipelined to produce
\begin{eqnarray}
 \frac{150\times 18750}{100\times 5000+50\times 18750}\approx 2 \nonumber
\end{eqnarray}
magic states in parallel. The physical qubits in the second layer is reused. Let $t_{sc} = 200 ns$ is a surface code cycle time. Then 2 magic states can be produced every $\sigma\times t_{sc} = 30\times 10^{-6}$ s. We require $\frac{\tcount}{\tcount_d} = \frac{5.4975\times 10^{11}}{14}\approx 3.9\times 10^{10}$ magic states per layer or depth of T-gates. We produce these many magic states in parallel for each layer. Due to parallelization the number of physical qubits required is
\begin{eqnarray}
 \frac{1}{2}\times 3.9\times 10^{10}\times 18750\approx 3.66\times 10^{14}
\end{eqnarray}
and the time taken is
\begin{eqnarray}
 14\times 30\times 10^{-6}s = 4.2\times 10^{-4}s.
\end{eqnarray}

In surface code implementation the cost of implementation of a multi-target CNOT is equal to the cost of a single target CNOT \cite{2012_FMMC} and has the same execution time. So we consider a multi-target CNOT as one logical CNOT. Each Toffoli can be implemented with 2 CNOTs, 2 multi-target CNOTs, 1 H and 1 S gate \cite{2018_G}. Thus we can upper bound the number of logical Cliffords by $7N = 7\times 2^{36} = 4.81\times 10^{11}$. The overall error rate of the Cliffords should therefore be less than $\frac{1}{7N} \approx 2.08\times 10^{-12}$. To compute the required distance, we seek the smallest $d$ that satisfies the inequality
\begin{eqnarray}
 \left( \frac{p_{in}}{0.0125} \right)^{\frac{d+1}{2}} < 2.08\times 10^{-12} \nonumber
\end{eqnarray}
and find this to be $d = 11$. The number of physical qubits required to encode the Cliffords is at most $2\times 2^{36}\times 2.5\times 1.25\times 11^2 \approx 5.197\times 10^{13}$. Overall, we require approximately $3.66\times 10^{14}+5.197\times 10^{13} \approx 4.18\times 10^{14}$ physical qubits to encode the complete $\qram_{poly}$. 

Roughly, we expect to perform $\frac{7N}{2N\times \tcount_d} = \frac{7}{2\times 14}\approx 0.25$ logical Clifford operations per qubit per layer of T-depth. Since each logical CNOT takes 2 surface code cycles and we require $\sigma = 150$ surface code cycles per T-depth, so the overall time is dominated by the time for implementing the non-Clifford T-gates. Thus, the time for implementing one memory read/write operation with $\qram_{poly}$ is approximately $4.2\times 10^{-4}$s and utilizes about $4.18\times 10^{14}$ physical qubits. 

The QRAM in \cite{2020_dMGM} requires $1.5\times 10^{15}$ physical qubits and is implemented in approximately $2.13\times 10^{-3}$s, using the same surface code parameters and doing a similar analysis. We remark that design considerations can vary. We can further reduce the time by parallelizing the magic state factories even more. This will increase the number of physical qubits.

\section{Other applications of polynomial encoding}
\label{sec:application}

In this section we discuss some applications for the polynomial encoding of bit-strings. First, we  describe a quantum look-up-table (qLUT) that can be built using two QRAMs. If we use $\qram_{poly}$ that uses the polynomial encoding, then our $\qlut_{poly}$ has double exponentially less T-depth than previous designs. Second, we descibe a method to optimize the Toffoli-count of circuits consisting of groups of multi-controlled-NOT gates.   

\subsection{Quantum look-up-table ($\qlut_{poly}$)}
\label{sec:qLUT}

\begin{figure}[!t]
 \centering
 \includegraphics[width = 13cm, height = 15cm]{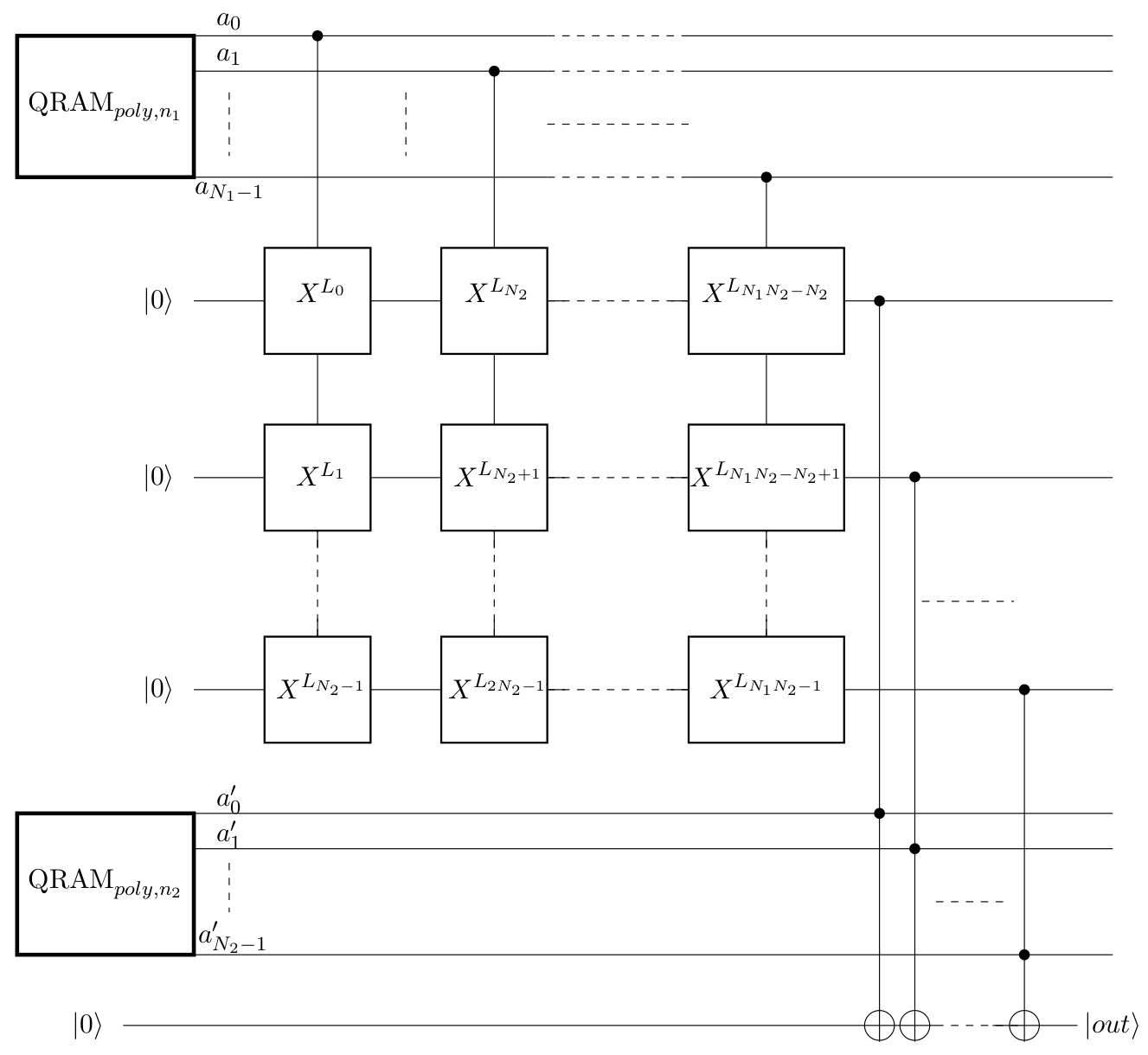}
 \caption{A quantum look-up table (qLUT), using $n_1$-bit and $n_2$-bit QRAM in parallel. For $\qlut_{poly}$ we use two $\qram_{poly}$, one addressing $n_1$-bit locations and another one addressing $n_2$-bit locations. }
 \label{fig:qlut}
\end{figure}

Now we describe a quantum look-up-table formed by combining two QRAMs, as shown in Figure \ref{fig:qlut}. Suppose we want to build a qLUT with $n$ address bits. That is, there are $N=2^n$ number of look-up addresses. We divide the address bits into two groups, the first one with $n_1$ bits and the next one with $n_2$ bits. That is, $n=n_1+n_2$ and assume $N_1=2^{n_1}$, $N_2=2^{n_2}$. With these we build two QRAMs, $\qram_{poly,n_1}$ and $\qram_{poly,n_2}$ with $n_1$ and $n_2$ number of address qubits respectively. Thus, $\qram_{poly,n_1}$ and $\qram_{poly,n_2}$ are capable of addressing $N_1$ and $N_2$ memory locations respectively. And as discussed before we have $N_1$ and $N_2$ number of (selecting) ancillae in each $\qram_{poly}$ (respectively) that correspond to these address locations. 

Suppose we want to read a look-up address indexed by the bit string $b_0,\ldots,b_{n-1}$, or alternatively by the integer $N' = \sum_{j=0}^{n-1}b_j2^j$. Now,
\begin{eqnarray}
 N' &=& \sum_{j=0}^{n-1}b_j2^j  = \sum_{j=0}^{n_2-1}b_j2^j+\sum_{j=n_2}^{n-1}b_j2^j 
 =\sum_{j=0}^{n_2-1}b_j2^j+2^{n_2}\left(\sum_{j=n_2}^{n-1}b_j2^{j-n_2}  \right)   \nonumber \\
 &=& \sum_{j=0}^{n_2-1}b_j2^j+N_2\left(\sum_{j'=0}^{n-n_2-1}b_{j'+n_2}2^{j'}  \right) =  \sum_{j=0}^{n_2-1}b_j2^j+N_2\left(\sum_{j'=0}^{n_1-1}b_{j'+n_2}2^{j'}  \right) = N_2'+N_2N_1'.     \nonumber
\end{eqnarray}
With the first QRAM i.e. $\qram_{poly,n_1}$ we select all address locations with first $n_1$ bits as $b_0,\ldots,b_{n_1-1}$. From the previous equation we know that there are $N_2$ such locations and we copy their contents in separate registers. With the next QRAM i.e. $\qram_{poly,n_2}$ we select an address among these copied addresses. Specifically, it selects an address with the last $n_2$ bits as $b_{n_1},\ldots,b_n$. Hence finally a memory location with address $(b_0,\ldots,b_n)$ gets selected. After this, using Toffolis we copy the contents of this memory location into the output bus.

Let each look-up address has $\ell$ bits. So after an address gets selected we require $N_2\ell$ Toffolis in order to compute the parity into the output bus. We can parallelize these Toffolis with additional $N_2\ell$ ancillae, as shown in Figure \ref{fig:Mpar} and explained in Section \ref{subsec:qramPar}. $\qram_{poly,n_1}$ and $\qram_{poly,n_2}$ can be implemented in parallel. Further, if we use the parallelized versions of these QRAMs then we have the following.
\begin{eqnarray}
 && \text{Toffoli-depth} = \max\{\log_2n_1,\log_2n_2\}+1   \nonumber \\
 && \text{Toffoli-count} =  (N_1-n_1-1)+(N_2-n_2-1)+\ell N_2 = N_1+(\ell+1)N_2-n-2 \nonumber \\
 && \#\text{Ancillae} = 2N_1+2N_2+\ell N_2+\ell N_2 = 2(N_1+(\ell+1)N_2) \nonumber
\end{eqnarray}
$2N_1$ and $2N_2$ ancillae are required to implement $\qram_{poly,n_1}$ and $\qram_{poly,n_2}$, respectively. $\ell N_2$ qubits are required to copy the contents of the subset of memory locations selected by $\qram_{poly,n_1}$. $\ell N_2$ qubits are also required to parallelize the last $\ell N_2$ Toffolis and compute the parity. 

\paragraph{Comparison with previous works :} In the CSWAP architecture for qLUT \cite{2024_LKS} the authors combine a QROM and a specific QRAM. The QROM is implemented with a set of multi-controlled-NOT gates. The QRAM is implemented with a number of Fredkin or controlled-SWAP unitaries. Each controlled-SWAP can be implemented with a Toffoli and CNOT. The contents of the selected memory location is always obtained on some specified qubits. This qLUT has a T-count of $O(\sqrt{N})$, T-depth $O(\sqrt{N})$ and number of qubits $O(\sqrt{N})$. 

If $N_1=N_2=\sqrt{N}$  then using any of the existing implementations of Toffoli, our $\qlut_{poly}$ has T-depth $O(\log_2\log_2N)$, which is a double exponential improvement over the previous work. The T-count and number of qubits is asymptotically same. Thus from a fault-tolerant perspective, assuming the cost metric in Equation \ref{eqn:qVol}, our design is expected to perform better. 

Our CNOT cost is primarily dominated by the step where we copy a subset of memory locations selected by the first QRAM. We require $\ell N_1N_2 = \ell N$ CNOTs at this stage. But again this is a group of $N_1$ multi-target CNOTs, where each has $N_2$ target. So in surface code implementation the execution time is equivalent to the time of execution of $N_1$ logical CNOTs. This cost is the same as that in \cite{2024_LKS}.

\subsection{Toffoli-count optimization of quantum circuits}
\label{subsec:TcountOpt}

The polynomial encoding can be used to optimize the number of Toffolis required to implement groups of multi-controlled-X gates or mixed polarity multiple control Toffolis (MPMCTs) \cite{2020_dMGM}. For example, we want to implement a circuit that flips a qubit to $\ket{1}$ for a subset, $\mathcal{S} \subseteq \{0,1\}^n$, of $n$-bit strings. These types of circuits also represent a kind of QROM. These can be used to select a subset of addresses and implement certain operations on those locations. We optimize the Toffoli-count of such circuits using the following procedure, which we call \textbf{TOFFOLI-OPT-POLY}.

\begin{figure}
\centering
\begin{subfigure}[b]{0.45\textwidth}
 \centering 
 \includegraphics[width=5cm, height=3cm]{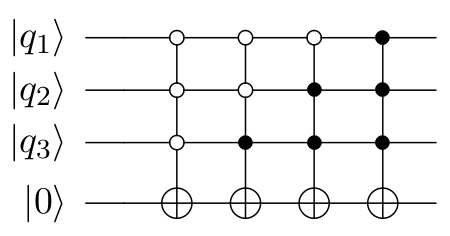}
 \caption{}
 \label{ckt:qrom3}
\end{subfigure}
\hfill
\begin{subfigure}[b]{0.45\textwidth}
 \centering 
 \includegraphics[width=5cm, height=3cm]{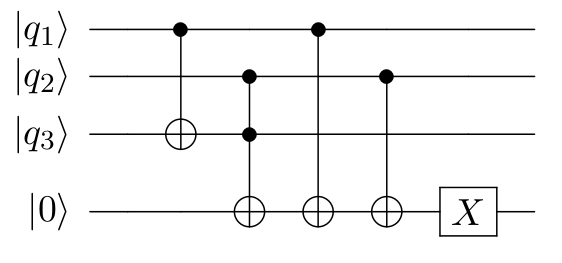}
 \caption{} 
 \label{ckt:qrom3Poly}
\end{subfigure}
 \caption{(a) A circuit with 4 multi-controlled-NOT gates or $C^3X$, where each such gate is controlled on 3 qubits. (b) The same circuit implemented with 1 Toffoli gate.  }
 \label{ckt:qrom}
\end{figure}

\begin{enumerate}
 \item Compute the encoding polynomial of each bit-string $\vec{b}\in\mathcal{S}$. This can be done conveniently using Lemma \ref{lem:monoSum}.
 
 \item Compute the following sum of the encoding polynomials.
 \begin{eqnarray}
  p(x_1,\ldots,x_n) = \bigoplus_{\vec{b}\in\mathcal{S}}p_{\vec{b}}(x_1,\ldots,x_n)  \nonumber
 \end{eqnarray}
 In this case coefficients of same monomials are added and reduced modulo 2.
 
 \item A product of linear polynomials can be implemented with a Toffoli. We remember that a linear polynomial is the sum of monomials of weight at most 1. Such a polynomial can be implemented with CNOTs and X gates. 
 
 Arrange the terms in $p(x_1,\ldots,x_n)$ such that the number of products of linear polynomials is optimized. 
 
 \item For each product we implement its factors in separate qubits, using CNOT and X gates. Using Toffoli we multiply these factors to implement the product. Then using CNOTs we add such product terms in order to implement $p(x_1,\ldots,x_n)$.
\end{enumerate}

For illustration, consider the circuit shown in Figure \ref{ckt:qrom3}, that has 3 qubits $q_1, q_2, q_3$ and another qubit initialized to $\ket{0}$. It flips the last qubit to $\ket{1}$ whenever the state of the first 3 qubits is $\ket{000}$, $\ket{001}$, $\ket{011}$ or $\ket{111}$. The encoding polynomials $p_{000}(x_1,x_2,x_3)$, $p_{001}(x_1,x_2,x_3)$, $p_{011}(x_1,x_2,x_3)$ and $p_{111}(x_1,x_2,x_3)$ have been calculated in Table \ref{tab:encodePoly3}. Then,
\begin{eqnarray}
 p(x_1,x_2,x_3) &=& p_{000}(x_1,x_2,x_3) \oplus p_{001}(x_1,x_2,x_3) \oplus p_{011}(x_1,x_2,x_3) \oplus p_{111}(x_1,x_2,x_3)    \nonumber \\
 &=& 1+x_1+x_2+x_1x_2+x_2x_3 = 1+x_1+x_2+x_2(x_1+x_3).  \nonumber
\end{eqnarray}
The 3 qubits $q_1, q_2, q_3$ are assigned variables $x_1, x_2, x_3$, respectively. With a CNOT controlled on $q_1$ and having target on $q_3$, we compute $x_1\oplus x_3$. Then using a Toffoli controlled on $\ket{q_2}$ and $\ket{q_3}$, that store $x_2$ and $x_1\oplus x_3$, respectively, we compute the product $x_2(x_1+x_3)$. The rest of the variables can be added using CNOTs and X. The optimized circuit with 1 Toffoli gate has been shown in Figure \ref{ckt:qrom3Poly}.

Here we observe that a sequence of multi-controlled-X gates represents a Boolean function which is a sum of product terms. We can find its ESOP (Exclusive Sum-of-Products) expression using tools like EXORCISM \cite{1993_SP, 1996_SP, 2001_AP}. Then factoring this expression we can implement the Boolean expression. We can also use the algorithm in \cite{2013_SSP} which first computes the ESOP, and then breaks the expression into common cofactors, which are reversibly synthesized. For example, in \cite{2020_dMGM} the authors mentioned that the circuit in Figure \ref{ckt:qrom3} can be implemented with 2 Toffolis. This is more than the Toffoli-count we get.
We can also implement each multi-controlled-NOT-X gate using the decomposition given in \cite{2017_HLZetal}. Each $C^nX$ i.e. an $n$-qubit-controlled-X gate can be implemented with $n-1$ Toffolis and an additional $n-1$ ancillae. But this will give more Toffoli-count than our implementation in Figure \ref{ckt:qrom3Poly}.

\section{Discussions and Conclusion}
\label{sec:discuss}

In this paper we develop a new design for quantum random access memory, using a polynomial encoding of the bit strings specifying the address of memory locations. 
 We implement a Clifford+T circuit for our $\qram_{poly}$ and 
show that $\qram_{poly} $ has T-count $O(N-\log N-1)$, T-depth $O(\log\log N)$ and uses $O(N)$ logical qubits. Thus with our design of $\qram_{poly}$ we achieve an exponential improvement in T-depth, while reducing T-count and keeping the number of logical qubits requirement the same with respect to the previous state-of-the-art bucket brigade architecture \cite{2008_GLM, 2020_dMGM}. 
 We illustrate that when encoded with the surface code \cite{2012_F, 2012_FMMC}, in order to perform one memory read/write operation, $\qram_{poly}$ takes less time and uses much less number of physical qubits. Using two such $\qram_{poly}$ we implement a quantum look-up table ($\qlut_{poly}$) that has T-count $O(\sqrt{N})$, T-depth $O(\log\log N)$ and uses $O(\sqrt{N})$ logical qubits. With our quantum look-up-table circuit $\qlut_{poly}$ we achieve (Table \ref{tab:compare}) a double exponential improvement in T-depth over the previous state-of-the-art CSWAP architecture for qLUT \cite{2024_LKS}, while the T-count and qubit-count are asymptotically same.

In our designs reduction in non-Clifford gate count comes at the cost of an increase in the CNOT gate count. The latter is a Clifford gate and in most error correction schemes the cost of implementing a Clifford is much less than the cost of implementing a non-Clifford. Thus, in our illustration we obtained better performance compared to previous QRAM designs. But CNOT, being a multi-qubit gate is more error prone than single-qubit gates like T. Even for connectivity constrained architectures (especially of the NISQ era) implementing a multi-qubit gate becomes more costly because it often needs a number of intermediate CNOT or SWAP gates, thus increasing the total gate count \cite{2022_GHLetal}. 

The problem of studying the noise-resilience of QRAM is an active research problem \cite{2015_AGJetal, 2021_HLGetal}, especially for the pre-fault-tolerant regime.
 And we believe it is beyond the scope of this current work because this paper exclusively focuses on performance improvements in the fault-tolerant regime. Often metrics and design considerations in these two regimes differ and hence they are studied separately. A detail analysis of the noise-resilience of $\qram_{poly}$ is left for future work. It will also be interesting to study the mapping overhead of these circuits in different architectural layouts like 2D grid, as done in \cite{2023_XHFetal, 2024_ZSL}. We expect to find trade-offs between the CNOT count and error rate or sparsity of the underlying graph. We can also aim at developing different hybrid designs with these new and existing circuits, so that we can take advantage of the various designs in different scenarios. 

Using the polynomial encoding, we develop a method (TOFFOLI-OPT-POLY) to optimize the Toffoli-count of quantum circuits, especially those using multi-controlled-NOT gates. Since such circuits represent a sum-of-product (SOP) form of Boolean function, so these  encodings can also have potential application in optimizing Boolean ESOP expressions, similar to the algorithms in \cite{1993_SP, 1996_SP, 2001_AP, 2013_SSP}. Some of these classical algorithms have inspired methods for reversible quantum logic synthesis \cite{2023_SBBetal}, which in turn have been an integral part of the design of quantum oracles for important algorithms like Grover's search. Thus these encoding polynomials may be used for reversible quantum logic synthesis. In the future we aim to investigate this avenue and the application of the polynomial encodings towards the design of algorithm-specific oracles and application-specific QRAMs, as in \cite{2019_PPR, 2022_NZBetal, 2023_PLG, 2023_SBBetal, 2024_NZ}.


\section*{Acknowledgement}

The author thanks Nathan Wiebe for helpful discussions. The author also thanks the anonymous reviewers whose helpful comments have helped us improve our manuscript significantly. The author acknowledges funding from the NSERC discovery program.

\section*{Author contributions}

The ideas, implementations and preparation of the manuscript was done by P.Mukhopadhyay.

\section*{Data availability}

All relevant data are included in this manuscript.

\section*{Competing interests}

There are no competing interests to declare.


\newcommand{\etalchar}[1]{$^{#1}$}

\end{document}